\documentclass[aps,prd,onecolumn,amsmath,amsfonts,amssymb,eqsecnum,groupedaddress,nofootinbib,nobalancelastpage,floatfix,superscriptaddress,secnumarabic,notitlepage,10pt]{revtex4-2}
\usepackage[utf8]{inputenc}
\usepackage{graphicx,multirow,xspace}
\usepackage[dvipsnames]{xcolor}
\usepackage{hyperref}
\hypersetup{%
  colorlinks=true,
  linkcolor=BurntOrange,
  citecolor=BurntOrange,
  filecolor=BurntOrange,
  urlcolor=BurntOrange
}
\usepackage{booktabs}
\usepackage{siunitx}
\DeclareUnicodeCharacter{2212}{-}

\newcommand{\be}{\begin{equation}}
\newcommand{\ee}{\end{equation}}
\newcommand{\bpm}{\begin{pmatrix}}
\newcommand{\epm}{\end{pmatrix}}

\newcommand{\ttbar}{\ensuremath{t\bar t}\xspace}

\DeclareMathOperator{\sign}{sgn}

\begin{document}
\title{Extracting a Toponium Signal at the LHC with Spin and Quantum Information Tools}

\author{Laura Antozzi}
\affiliation{University of Bologna, Department of Physics and Astronomy, Italy}

\author{Esteban Chalbaud}
\affiliation{Laboratório de Instrumentação e Física Experimental de Partículas, Department of Physics, University of Coimbra, 3004-516 Coimbra, Portugal}

\author{Frédéric Déliot}
\affiliation{Université Paris-Saclay, CEA, Département de Physique des Particules, 91191, Gif-sur-Yvette, France}

\author{Federica Fabbri}
\affiliation{Istituto Nazionale di Fisica Nucleare, Sezione di Bologna, Italy}
\affiliation{University of Bologna, Department of Physics and Astronomy, Italy}

\author{Miguel~C.N.~Fiolhais}
\affiliation{Laboratório de Instrumentação e Física Experimental de Partículas, Department of Physics, University of Coimbra, 3004-516 Coimbra, Portugal}
\affiliation{Science Department, Borough of Manhattan Community College, The City University of New York, 199 Chambers St, New York, NY 10007, USA}
\affiliation{The Graduate School and University Center, The City University of New York, 365 Fifth Avenue, New York, New York 10016, USA}

\author{Benjamin Fuks}
\affiliation{Laboratoire de Physique Théorique et Hautes Énergies (LPTHE), UMR 7589, Sorbonne Université et CNRS, 4 place Jussieu, 75252 Paris Cedex 05, France}

\author{António Onofre}
\affiliation{Laboratório de Instrumentação e Física Experimental de Partículas, Department of Physics, University of Coimbra, 3004-516 Coimbra, Portugal}
\affiliation{ Centro de Física da Universidade do Minho e Universidade do Porto (CF-UM-UP), Universidade do Minho, 4710-057 Braga, Portugal}

\author{Martin White}
\affiliation{ARC Centre of Excellence for Dark Matter Particle Physics \& CSSM, Department of Physics, Adelaide University, Adelaide, SA 5005, Australia}

\author{Pengxuan Zhu}
\affiliation{ARC Centre of Excellence for Dark Matter Particle Physics \& CSSM, Department of Physics, Adelaide University, Adelaide, SA 5005, Australia}

\date{\today}
\begin{abstract}
  We investigate near-threshold top-antitop production at the LHC, focusing on the impact of toponium formation on spin correlations and quantum information properties of the final state. Considering the top-antitop system as a mixed two-qubit state, we reconstruct spin density matrices via quantum tomography and evaluate several observables including some inspired by quantum information. We then compare their sensitivity in discriminating toponium effects from top-antitop production without these effects. Our results demonstrate that combining these variables is expected to significantly enhance sensitivity to toponium effects, bringing new ways to explore these subtle features.  
\end{abstract}

\maketitle

\section{Introduction}

The top quark occupies a singular position in the Standard Model (SM) of particle physics. As the heaviest known elementary particle with a Yukawa coupling to the Higgs field of order unity, it is uniquely sensitive to the electroweak symmetry-breaking mechanism and is widely believed to be well-connected to yet-unknown new dynamics beyond the SM~\cite{Bernreuther:2008ju, Franceschini:2023nlp}. In particular, the extremely short lifetime of the top quark prevents it from hadronising, thereby allowing it to transmit information about its quantum properties directly to its decay products. 
This feature has made top-quark pair (\ttbar) production via the strong interaction an important laboratory for precision studies of QCD at colliders, for searches for new interactions and increasingly for the application of quantum information-inspired observables at the LHC. 
This also offers a unique opportunity to study the properties of a bare quark and, in particular, its spin properties. 
While top quarks are primarily produced in unpolarised pairs, the top and antitop spins are indeed still correlated with a strength depending on the spin quantisation axis.

In this context, would-be bound states of a top-antitop system, generically referred to as toponium, were predicted~\cite{Fadin:1987wz, Fadin:1988fn} long before the top quark discovery in 1995~\cite{CDF:1995wbb, D0:1995jca}, and believed to be unobservable because the weak decay of the top quark occurs on a timescale shorter than needed for non-perturbative binding to fully develop. Nevertheless, it has been recognised that near-threshold dynamics remain governed by non-relativistic QCD, and remnants of the bound-state behaviour such as threshold enhancements and quasi-resonant structures can persist even in the presence of the top quark's rapid decay~\cite{Fadin:1990wx, Strassler:1990nw, Jezabek:1992np, Sumino:1992ai, Hagiwara:2008df, Sumino:2010bv}. Despite these challenges, recent experimental indications of anomalous structures in the $t\bar t$ dilepton invariant-mass and angular spectra~\cite{ATLAS:2023fsd, CMS:2024pts, CMS:2024ybg, CMS:2024zkc, CMS:2025kzt, CMS:2025dzq, ATLAS:2026dbe} have renewed interest in toponium-like phenomena. In an early study of partial run~2 data~\cite{Fuks:2021xje}, it was indeed emphasised that the observed excesses could be compatible with toponium formation, which triggered numerous new analyses exploring this possibility, sometimes even in connection with physics beyond the SM~\cite{Maltoni:2024tul, Maltoni:2024wyh, Aguilar-Saavedra:2024mnm, Lu:2024twj, Fuks:2024yjj, Llanes-Estrada:2024phk, Djouadi:2024lyv, Garzelli:2024uhe, Francener:2025tor, Nason:2025hix, Fuks:2025sxu, Shao:2025dzw, Fuks:2025wtq, Fuks:2025toq, Flacke:2025dwk}. In particular, a practical strategy to incorporate toponium effects into the Monte Carlo event generators routinely employed in LHC analyses has recently been developed~\cite{Fuks:2024yjj, Fuks:2025wtq, Flacke:2025dwk}. The method exploits the Green's functions of the non-relativistic QCD Hamiltonian~\cite{Fadin:1987wz, Fadin:1990wx, Strassler:1990nw, Jezabek:1992np, Sumino:1992ai, Hagiwara:2008df} to encode the bound-state dynamics and the $t\bar t$ threshold enhancements, and uses them to re-weight the conventional $t\bar t$ production and decay matrix elements. This subsequently enables particle-level predictions for a broad class of LHC observables while systematically including Coulombic and quasi-bound-state effects. In particular, since the toponium dynamics modify the quantum state of the $t\bar t$ system near threshold, they are expected to leave characteristic imprints in spin correlations and angular observables, which are directly sensitive to the underlying quantum coherence of the pair.

The correlations between the spins of a pair of top quarks produced at the LHC can indeed be interpreted as manifestations of the quantum correlations inherent to the scattering process.
The natural language to describe such correlations is provided by the theory of quantum information which, for bipartite systems, offers various metrics that characterise the information content of the correlations between the two subsystems. Recent years have seen an explosion of activity in applying quantum information concepts to the study of high energy physics systems, including mapping the quantum information theoretic content of particle scattering processes, studying connections between quantum information principles and emergent symmetry in the SM and its extensions, and measuring quantum information properties at the LHC~\cite{Balasubramanian:2011wt, Seki:2014cgq, Peschanski:2016hgk, Grignani:2016igg, Kharzeev:2017qzs, Fan:2017hcd, Fan:2017mth, Cervera-Lierta:2017tdt, Carena:2023vjc, Low:2024hvn, Carena:2025wyh, Liu:2025pny, Hu:2025lua, Sou:2025tyf, Afik:2020onf, Aoude:2022imd, Fabbrichesi:2021npl, Severi:2021cnj, Afik:2022kwm, Aguilar-Saavedra:2022uye, Fabbrichesi:2022ovb, Afik:2022dgh, Aguilar-Saavedra:2023hss, Han:2023fci, Aguilar-Saavedra:2024hwd, ATLAS:2023jzs, ATLAS:2023fsd, CMS:2024hgo, Abel:1992kz, White:2024nuc, Liu:2025frx, Liu:2025qfl, CMS:2025cim, Aoude:2025jzc, Barr:2022wyq, AshbyPickering:2022umy, Aguilar-Saavedra:2022wam, Aguilar-Saavedra:2022mpg, Fabbrichesi:2023jep, Fabbrichesi:2023idl, Fabbri:2023ncz, Aoude:2023hxv, Fabbrichesi:2023cev, Dong:2023xiw, Altomonte:2023mug, Aguilar-Saavedra:2024vpd, Aguilar-Saavedra:2024whi, Morales:2024jhj, Fabbrichesi:2024rec, LoChiatto:2024dmx, Han:2024ugl, Horodecki:2025tpn, DelGratta:2025qyp, Nason:2025hix, Grossi:2024jae, Cheng:2025cuv, Ghodrati:2023uef, Gargalionis:2025iqs, Fabbrichesi:2025psr, Goncalves:2025mvl, Goncalves:2025xer}. In particular, a comprehensive framework for calculating density matrices from simulated collider events and using them to perform quantum information tests was recently presented in Ref.~\cite{Durupt:2025wuk}, whilst two detailed recent reviews of the use of quantum information techniques at particle colliders can be found in Refs.~\cite{Barr:2024djo,Afik:2025ejh}. 

As noted above, toponium effects inevitably change the structure of the spin correlations in \ttbar\  events at the LHC, which means that it also changes the quantum information structure in the final state. This state is a mixed state, as opposed to a pure state due to imperfect knowledge of the incoming state at the LHC, it is nevertheless possible to define and measure quantum-information theoretic quantities to characterise it. These should indeed show different behaviour for \ttbar\  events in the absence of toponium and \ttbar\  events in the presence of toponium. Furthermore, detailed study of relevant quantum information properties may allow us to better characterise the nature of toponium formation. For these reasons, in this work we explore a variety of quantum information measures through a study of simulated \ttbar\  events, comparing and contrasting their discriminating power with that of standard kinematic observables. This also deals with an open question that was previously partially addressed in Ref.~\cite{Aoude:2025jzc}, namely if quantum information observables are useful or redundant when probing novel physics effects. This is expected to be physics-process dependent, but toponium dynamics serves as a case study that is as topical as it is useful and informative.

This paper is structured as follows. In Section~\ref{sec:spin}, we review the necessary formalism for performing spin correlation measurements and quantum information tests from the reconstructed four vectors of particles  at the LHC. Section~\ref{sec:sim} provides the details of our Monte Carlo simulation chain, before we present and discuss our results in Section~\ref{sec:analysis}. We provide our final conclusions in Section~\ref{sec:conclusion}.

\section{Spin and Quantum Information}
\label{sec:spin}

In this section, we briefly review the quantum information concepts relevant to this work, largely following Refs.~\cite{Barr:2024djo,Durupt:2025wuk} in which more details can be found. The spin-1/2 nature of the top quark means that, from a quantum information perspective, a \ttbar\ pair acts like a two-qubit system. Labelling the two top quarks as $A$ and $B$, we can represent their individual states in the Hilbert spaces $\mathcal{H}_A$ and $\mathcal{H}_B$, and the total quantum state in the Hilbert space $\mathcal{H}=\mathcal{H}_A\otimes\mathcal{H}_B$. Top-quark pair production at the LHC occurs via the strong interaction with contributions arising from different partonic initial states, primarily gluon-gluon fusion and quark-antiquark annihilation. These contributions enter the cross section through an incoherent sum, weighted by the corresponding parton distribution functions. As a result, the produced $t\bar t$ system is not described by a pure quantum state, but rather by a statistical mixture with probabilities determined by the relative contributions of the different production channels and the final-state kinematics. The standard way of dealing with such mixed states is to define a density matrix $\rho$ as
\begin{equation}
\rho = \sum_i p_i|\psi_i\rangle\langle \psi_i|,
\end{equation}
where $p_i$ is the probability that the system is parametrised by any particular state $|\psi_i\rangle$. For a mixed state, $\rho$ satisfies 
\begin{equation}
  z\rho\ge 0,~~~\text{Tr}\left[\rho\right]=1,~~~0\le\text{Tr}\left[\rho^2\right]\le 1.
\end{equation}
Assuming that $\rho$ is the full density matrix that operates on the Hilbert space $\mathcal{H}$, we can also define the reduced density matrices for the subsystems $A$ and $B$ via
\begin{equation}
  \rho_A=\text{Tr}_B\left[\rho\right]
  \qquad\text{and}\qquad
  \rho_B=\text{Tr}_A\left[\rho\right],
\end{equation}
where $\text{Tr}_X$ denotes a partial trace over the subsystem $X\in\{A,B\}$. A state is called separable if the density matrix has the special form
\begin{equation}
\rho=\sum_i p_i\, \rho_A^{(i)} \otimes \rho_B^{(i)}, 
\end{equation}
with $0\le p_i \le 1$. Otherwise it is said to be entangled.

To make this more concrete, we consider the density matrix associated with the \ttbar\  system. This is a $4\times 4$ matrix that can be parametrised in various ways. The most popular one is the Pauli or Fano-Bloch expansion, given by
\begin{equation}
\label{eq:fano}
\rho = \frac{1}{4}
\left(
\, \mathbb{I}_4
+ \sum_i B_{1i}\, (\sigma_i \otimes \mathbb{I}_2)
+ \sum_j B_{2j}\, (\mathbb{I}_2 \otimes \sigma_j)
+ \sum_{i,j} C_{ij}\, (\sigma_i \otimes \sigma_j)
\right).
\end{equation}
In this expression, we have denoted the $n$-dimensional identity matrix as $\mathbb{I}_n$ and the Pauli matrices as $\sigma_i$ (where $i\in \{1,2,3\}$). The vectors $B_1$ and $B_2$ represent the single-particle polarisation vectors, whilst the $C_{ij}$ coefficients are the elements of the $3\times 3$ spin-correlation matrix. These various parameters, also known as the Fano coefficients, can be reconstructed via the tool of quantum state tomography that is discussed in section~\ref{sec:observables}. We assume in the following that the indices $i$ and $j$ run over the axes of the helicity basis $\{\hat k,\hat r,\hat n\}$. Although several choices are possible for this reference system suitable to analyse spin effects, our choice is based on the decomposition of the $t\bar t$ production spin-density matrix on an orthonormal basis defined in the $t\bar t$ rest frame~\cite{Aguilar-Saavedra:2022uye}. In this construction, the unit vectors are specified by defining $\hat k$ to be aligned with the top-quark momentum in the $t\bar t$ rest frame, fixing $\hat p$ as the incoming beam direction and defining the remaining unit vectors as
\begin{equation}
\label{eq:rkn_refsys}
\hat r = \frac{1}{r}(\hat p - y \hat k) \quad\text{and}\quad
\hat n = (\hat k \times \hat r),
\qquad\text{with}\qquad 
y = \hat k \cdot \hat p                 \quad\text{and}\quad
r = \sign(y)\sqrt{1-y^2}.
\end{equation}

Various quantum information metrics can be computed as functions of $\rho$, and the form in Eq.~(\ref{eq:fano}) allows us to express these in terms of the Fano coefficients. In this study, we compare and contrast observables including purity, $D$ coefficients and concurrence, logarithmic negativity, trace distance and magic. Purity measures how mixed a quantum state is, which means it should always be non-zero for \ttbar\  events. If extra processes are present in an event sample, however, we would expect the state to appear more mixed, thus giving potentially measurable differences in purity. In its simplest form, purity is defined as
$\gamma(\rho) = \mathrm{Tr}\,\left[\rho^{2}\right]$, but we instead use the so-called normalised purity which is bounded between zero and one and can be expressed, for a two-qubit system, as 
\begin{equation}
\mu(\rho) = \frac{4}{3}\!\left(\gamma(\rho) - \frac{1}{4}\right).
\end{equation}
The upper bound of $\mu(\rho)=1$ is only reached for pure states, and should thus not be observed in LHC measurements. The lower bound is instead reached for $\rho=\mathbb{I}_4/4$, which represents a maximally-mixed state. In the helicity basis, we have
\begin{equation}
\mu(\rho) =  \sum_{i\in \{k,r,n \}}\frac{B_{1i}^2  +  B_{2i}^2}{3} \ +\  \sum_{i,j\in \{k,r,n\}} \frac{C_{ij}^2}{3}.
\end{equation}

Another popular set of observables for studying angular correlations in top events at hadron colliders are the so-called $D$ coefficients~\cite{PhysRevD.53.4886}, given by the following functions of the Fano coefficients,
\begin{equation}\begin{split}
D^{(1)} =&\ \frac{1}{3}\,(C_{11} + C_{22} + C_{33}),  \\
D^{(x)} =&\ \frac{1}{3}\,(C_{11} - C_{22} - C_{33}),  \\
D^{(y)} =&\ \frac{1}{3}\,(-C_{11} + C_{22} - C_{33}), \\
D^{(z)} =&\ \frac{1}{3}\,(-C_{11} - C_{22} + C_{33}).
\end{split}\label{eq:d1}\end{equation}
In these expressions, $\{x,y,z\}$ label the coordinate axes of any particular spatial basis. The $D^{(i)}$ coefficients (with $i=1,x,y,z$) can be used to construct the concurrence, which serves as an entanglement witness and is defined by
\begin{equation}
\mathcal{C} = \frac{1}{2}\,\max\bigl(0,\,-1 - 3D_{\min}\bigr),
\end{equation}
with $D_{\min}$ defined as the smallest of the $D^{(i)}$ values. For a two-qubit system, the concurrence satisfies $0 \le \mathcal{C} \le 1$, with the lower bound signifying separability and the upper bound signifying a maximally-entangled state. The concurrence becomes positive and non-zero if any of the $D^{(i)}$ is smaller than $-1/3$, which provides an additional test of the entanglement of the top spins. Another widely used entanglement measure is the logarithmic negativity~\cite{Plenio:2007zz}, 
\begin{equation}
E_N(\rho) \equiv \log_2 \left(||\rho^{T_B}||\right).
\label{eq:e_n}
\end{equation}
Here, $\rho^{T_B}$ denotes the partial transpose of the density matrix with respect to the subsystem $B$, defined as $\rho^{T_B} \equiv (\mathbb{I}_2 \otimes \mathrm{T}) \rho$ in which $\mathrm{T}$ represents standard matrix transposition. Furthermore, we remind that the trace norm of a matrix $X$ is defined as
\begin{equation}
||X|| \equiv \operatorname{Tr}\left( \sqrt{X^\dagger X}\,\right).
\end{equation}
Since $\rho^{\mathrm{T}_B}$ is Hermitian, its trace norm is equal to the sum of the absolute values of its eigenvalues. The larger is the value of $E_N$, the greater is the entanglement of the two top spins. For the fully-general expansion of $\rho$ given in Eq.~(\ref{eq:fano}), it is not possible to write a simple, closed-form expression for $E_N$ in terms of the Fano coefficients, but it is straightforward to evaluate numerically. As we shall see in Section~\ref{sec:results}, however, in the case of top pair production at the LHC, it is a good approximation to assume that the only non-zero Fano coefficients are the $C_{kk}$, $C_{nn}$ and $C_{rr}$ ones. In this case, it can be shown that
\begin{equation}
E_N(\rho) = \frac{1}{8}\Big(
-4
+\bigl|1-C_{kk}-\Delta_-\bigr|
+\bigl|1-C_{kk}+\Delta_-\bigr|
+\bigl|1+C_{kk}-\Delta_+\bigr|
+\bigl|1+C_{kk}+\Delta_+\bigr|
\Big),
\end{equation}
where $\Delta_\pm \equiv \bigl(C_{nn}\pm C_{rr}\bigr)\,$. In our study, we however rely on the full formula of Eq.~\eqref{eq:e_n}.

The trace distance is used to compare two different density matrices, and it represents in some sense the distance between them taken as a trace norm over the difference between the matrices. A way to apply this for toponium studies is to consider first a prediction of the \ttbar\ density matrix $\rho_{\mathrm{sm}}$ in the absence of toponium effects (despite the fact that strictly speaking, toponium contributions are part of the SM prediction), allowing the trace distance for the measured density matrix $\rho$ to be defined as
\begin{equation}
D_T(\rho)
\equiv
\frac{1}{2}\text{Tr}
\left[
\sqrt{(\rho - \rho_{\mathrm{sm}})^{\dagger}(\rho - \rho_{\mathrm{sm}})}
\right].
\label{eq:d_t}
\end{equation}
As in the case of the logarithmic negativity, the trace norm in this expression does not permit a convenient closed-form expression in terms of the Fano coefficients in the most general case, but numerical evaluation is straightforward. In the case where only $C_{kk}$, $C_{nn}$ and $C_{rr}$ are non-zero, it can be shown that
\begin{equation}
\begin{split}
D_T(\rho) = &\ \frac{1}{8}\Big(
\bigl|\, (C_{kk}+C_{nn}-C_{rr})-(C_{kk}^{\rm sm}+C_{nn}^{\rm sm}-C_{rr}^{\rm sm}) \,\bigr|
+\bigl|\, (C_{kk}-C_{nn}+C_{rr})-(C_{kk}^{\rm sm}-C_{nn}^{\rm sm}+C_{rr}^{\rm sm}) \,\bigr| \\
&\quad +\bigl|\, (-C_{kk}\!+\!C_{nn}\!+\!C_{rr})-(-C_{kk}^{\rm sm}\!+\!C_{nn}^{\rm sm}\!+\!C_{rr}^{\rm sm}) \,\bigr|
+\bigl|\, (C_{kk}\!+\!C_{nn}\!+\!C_{rr})\!-\!(C_{kk}^{\rm sm}\!+\!C_{nn}^{\rm sm}\!+\!C_{rr}^{\rm sm}) \,\bigr|
\Big).
\end{split}
\end{equation}
However, as for the logarithmic negativity, our analysis relies on the full formula of Eq.~\eqref{eq:d_t}.

Finally, in quantum computing, certain states known as stabiliser states offer no advantage for quantum computation over classical computation. Interestingly, some stabiliser states are maximally-entangled Bell states, demonstrating that superposition and entanglement alone are not sufficient for a quantum advantage. This observation has motivated the definition of quantities that measure the deviation from stabiliser states, which is conventionally referred to as magic in the quantum information literature. Magic was first studied in the context of \ttbar\  production in Ref.~\cite{White:2024nuc}, which utilised the Second Stabiliser R\'enyi entropy $M_2$~\cite{Leone:2021rzd}. In terms of the full set of Fano coefficients, it is given by
\begin{equation}\renewcommand{\arraystretch}{1.5}
\label{eqn:magic}
M_2=-\log_2 \left(\frac{
1 + \sum_{i\in \{k,r,n\}} \left[B_{1i}^4 +B_{2i}^4\right]+\sum_{i,j\in \{k,r,n\}}C_{ij}^4}{
1 + \sum_{i\in \{k,r,n\}} \left[B_{1i}^2 +B_{2i}^2\right]+\sum_{i,j\in \{k,r,n\}}C_{ij}^2
}\right).
\end{equation}

\section{Simulation Toolchain}
\label{sec:sim}

The performance of the quantum-information-inspired observables and spin-correlation variables introduced in the previous section is evaluated using simulated event samples for standard \ttbar\ production (`background') and toponium-enhanced contributions (`signal'). Here, the term `signal' refers to $t\bar t$ events in which the intermediate top-antitop pair is produced in a colour-singlet configuration and undergoes bound-state effects, while the term `background' denotes the conventional perturbative predictions for the same final state. The \ttbar\ production process is simulated at next-to-leading order (NLO) in QCD using the Monte Carlo event generator \textsc{Powheg Box v2}~\cite{Frixione:2007nw, Nason:2004rx, Frixione:2007vw, Alioli:2010xd}. The hard-scattering matrix elements are convoluted with the NLO set of \textsc{NNPDF3.0} parton distribution functions~\cite{NNPDF:2014otw}, and the renormalisation and factorisation scales are set to the average transverse mass of the final-state particles. The events generated in this way are then interfaced with \textsc{Pythia 8.230}~\cite{Sjostrand:2014zea} for parton showering and hadronisation, using the standard A14 ATLAS tune~\cite{TheATLAScollaboration:2014rfk}. However, in our exploratory study, the analysis is performed at parton level such that hadronisation effects are neglected in order to isolate the impact of the considered observables on the underlying partonic dynamics. A more realistic treatment including full hadronisation and the simulation of the response of a typical LHC detector is left for future work.

Toponium effects are simulated using Monte Carlo event generation combined with a non-relativistic QCD framework tailored to the near-threshold regime of top-antitop production. Following the strategy outlined in~\cite{Fuks:2024yjj, Fuks:2025wtq, Flacke:2025dwk}, the short-distance production and decay of the $t\bar t$ system are described using standard relativistic $2\to6$ matrix elements, computed with \textsc{MG5\_aMC}~\cite{Alwall:2014hca}. In addition, the long-distance Coulombic interactions are incorporated through the S-wave Green's function of the non-relativistic QCD Hamiltonian~\cite{Fadin:1987wz, Fadin:1990wx, Hagiwara:2008df, Sumino:2010bv}, computed using a tree-level Coulomb potential. This Green's function encodes the resummation of ladder gluon exchanges and provides the leading contributions to the quasi-bound-state dynamics, and the effects are implemented  via the re-weighting of conventional $t\bar t$ events with weights constructed from the ratio of the interacting Green's function to the free Green's function. Moreover, toponium simulations are restricted to the near-threshold region where non-relativistic dynamics dominate, and parton-level events are passed to \textsc{Pythia} for parton showering and hadronisation. In this way, full spin correlations and decay kinematics are preserved at the particle level, enabling predictions for quantum information observables while consistently incorporating toponium-induced distortions of the $t\bar t$ phase space.

The reconstruction of the whole spin density matrix relies on quantum tomography~\cite{Afik:2020onf}. This means that the decay products' directions are exploited to reconstruct the spin of the parent particle, and that the quantum state is extracted by averaging. We only focus on the process
\begin{equation}
  t\bar{t}\rightarrow WbW\bar{b}\rightarrow \ell^+\nu_l\ell^-\bar{\nu_l} b \bar{b},
\end{equation}
and two different event selections are considered. The first is a minimal selection in which no kinematic requirements are imposed beyond restricting the phase space to $m_{t\bar t} < 355$~GeV. Here, the $t\bar t$ invariant mass is constructed from the four-momenta of the top quarks prior to their decay and after parton showering. This requirement then selects the threshold region for top-quark pair production and defines the only region in which bound-state dynamics are expected to significantly impact the predictions. The second selection is designed to approximate the requirements of a realistic dilepton analysis at the LHC. All lepton candidates are required to satisfy $|\eta| < 2.5$, and events must contain at least one lepton with $p_{\mathrm T} > 25$~GeV, thus emulating the single-lepton trigger requirement used during the 2022-2025 data-taking period by the ATLAS experiment~\cite{ATLAS:2008xda}. An alternative requirement, demanding at least two leptons with $p_{\mathrm T} > 20$~GeV, was also investigated. This resulted in a slightly reduced event yield but did not lead to significant differences in the distributions of the observables considered, and was therefore not pursued further. In addition, events are required to contain at least two $b$-quarks with $p_{\mathrm T} > 25$~GeV and $|\eta| < 2.5$. This reflects the need to exploit jets originating from $b$-quarks to isolate the $t\bar t$ topology and emulates the restriction of $b$-tagging to the central detector region. Finally, all observables are constructed directly from the generator-level top quarks rather than from the reconstructed final-state objects. The association of leptons to their parent top quarks is therefore unambiguous and fully determined by the lepton charge.

The observables constructed from truth-level top-quark information are combined into a multivariate discriminant based on Boosted Decision Trees (BDTs) to enhance the separation between the toponium signal and the \ttbar\ background. The classifiers are implemented using the \textsc{XGBoost} framework~\cite{Chen:2016btl, Cornell:2021gut} with the \texttt{gbtree} booster option. Several configurations are considered, differing in their choice of input variables: one using the full set of observables collected in Section~\ref{sec:observables}, one excluding the quantum-information-inspired variables, one excluding $p^\star$ (defined as the magnitude of the top-quark three-momentum in the \ttbar\ rest frame), and one restricted to the variables $D^{(1)}$ and the angle $\theta'$ between the positively charged lepton and the reflected momentum of the negatively charged lepton in the rest frame of their parent top quark (see Eq.~\eqref{eq:Dn-costheta} below). For configurations including the $p^\star$ quantity, a two-stage training strategy is adopted, since the simulated toponium $p^\star$ spectrum is only reliable in phase-space regions where the top quarks are non-relativistic without the implementation of a more accurate matching prescription~\cite{Fuks:2024yjj, Fuks:2025wtq}. Events with $p^\star < 35$~GeV are hence evaluated using a BDT trained with $p^\star$ included as an input variable while events with $p^\star \geq 35$~GeV are evaluated using a separate BDT trained without $p^\star$, and the final discriminant score is taken from the corresponding classifier according to the event’s $p^\star$ value. In this process, all simulated events are randomly divided into equal training and testing samples; no significant overtraining is observed, and variations of the BDT hyper-parameters have a negligible impact on the analysis performance.

\section{Probing Toponium Effects with Quantum-Information Observables}\label{sec:analysis}
To investigate the subtle influence of bound-state dynamics in top-quark pair production, we develop a multivariate analysis framework that combines conventional kinematic observables with quantum-information-inspired variables sensitive to spin correlations. In Section~\ref{sec:observables}, we first introduce the full set of observables used as inputs to the BDT classifiers. Section~\ref{sec:results} then presents the performance of the classifiers under different input configurations, quantifying the discriminating power between the toponium signal and the conventional \ttbar\ background and highlighting the relative contributions of the various classes of observables.

\subsection{Input Observables}
\label{sec:observables}

Toponium effects modify the spin correlations between the top and antitop quarks produced at the LHC, leading to characteristic distortions in the angular distributions of their decay products. In particular, the charged leptons from the intermediate $W$-boson decays inherit information about the top-quark spins, making leptonic angular observables especially sensitive to the bound-state dynamics. Near the production threshold, these are indeed expected to strengthen the correlations along specific spin axes, producing measurable deviations from the patterns predicted by conventional \ttbar\ production. Suitably chosen angular observables have been therefore extensively studied at the LHC, where dileptonic and lepton+jets final states have been measured with sufficient precision to test SM predictions and to search for deviations that could signal new effects in the top-quark sector~\cite{ATLAS:2019zrq,CMS:2024zkc,CMS:2019nrx}.

The spin correlations of the top-antitop system are conveniently analysed in a right-handed orthonormal basis defined by the axes $\{\hat k, \hat r, \hat n\}$. In this basis, the spin-dependent part of the production density matrix can be decomposed into coefficients associated with the correlations along these axes, and each coefficient can be classified according to its transformation properties under discrete symmetries such as parity ($P$), charge conjugation and parity ($CP$), naive time reversal ($T_N$) and $CPT_N$. This classification subsequently allows for the construction of angular observables that are sensitive to specific components of the spin correlation matrix while remaining insensitive to others, providing a clean way to separate different contributions to \ttbar\ production at the LHC~\cite{Bernreuther:2015yna_CP}. In addition, these observables are specifically interesting to study because their role extends beyond precision tests of the SM~\cite{PhysRevD.95.095012, PhysRevLett.76.4468}. In particular, they have been shown to enhance the sensitivity to heavy neutral Higgs bosons produced in association with, or decaying into, $t\bar t$ pairs. In scenarios such as type-II two-Higgs-doublet models, heavy scalar or pseudoscalar resonances can indeed contribute to top-antitop production and interfere with the non-resonant QCD continuum, leading to distortions in the invariant mass spectrum and modifications of spin-dependent observables. These effects then allow the study of angular observables to discriminate between scalar and pseudoscalar exchanges, whether measured in the laboratory frame or in the $t\bar t$ rest frame~\cite{Gunion:1996vv, Azevedo:2022jnd}. In our analysis, we consider the following CP-sensitive normalised operators,\footnote{We have verified that, among all variables proposed in Refs.~\cite{Gunion:1996vv, Azevedo:2022jnd}, $a_1 = \frac{(\vec p_t \times \hat n)\cdot(\vec p_{\bar t} \times \hat p)}{\left|(\vec p_t \times \hat p)\cdot(\vec p_{\bar t} \times \hat p)\right|}$ and $a_2 = \frac{p_{x,t} \, p_{x,\bar t}}{\left|p_{x,t} \, p_{x,\bar t}\right|}$ do not exhibit significant discriminating power and are therefore not considered further in this analysis.}
\begin{equation}
\label{eqn:b1tob4}
b_1 = \frac{(\vec p_t \times \hat p)\cdot(\vec p_{\bar t} \times \hat p)}
{p_{\mathrm{T},t} \, p_{\mathrm{T},\bar t}},\qquad
b_2 = \frac{(\vec p_t \times \hat n)\cdot(\vec p_{\bar t} \times \hat n)}
{|\vec p_t| \, |\vec p_{\bar t}|}, \qquad
b_3 = \frac{p_{x,t} \, p_{x,\bar t}}
{p_{\mathrm{T},t} \, p_{\mathrm{T},\bar t}}, \qquad
b_4 = \frac{p_{z,t} \, p_{z,\bar t}}
{|\vec p_t| \, |\vec p_{\bar t}|},
\end{equation}
where we remind that $\hat p$ is the unit vector along the beam direction and emphasise that all four-momenta are evaluated in the laboratory frame. Moreover, $p_{\mathrm{T},t}$ and $p_{\mathrm{T},t}$ denote the transverse momentum of the top and antitop quarks, respectively, whereas $p_{x,t}$ and $p_{x,\bar t}$ ($p_{z,t}$ and $p_{z,\bar t}$) refer to their projection in the $\hat x$ ($\hat z$) direction.

As discussed above, collider experiments do not have dedicated detectors for directly measuring the particle spins, so the spin information must be inferred from measurable quantities such as the kinematic distributions of the final-state particles. Thanks to the chiral nature of the weak interactions, which correlates the parent particle spin with the direction of its decay products, the spin density matrix of the \ttbar\ system can be reconstructed by averaging the angular distributions of the decay products over many events. The coefficients of the spin density matrix introduced in Eq.~(\ref{eq:fano}) can then be directly related to the differential cross section for top-quark pair production,
\begin{equation}
\label{QT}
\frac{1}{\sigma}
\frac{\mathrm{d}^4\sigma}{\mathrm{d}\Omega_1\, \mathrm{d}\Omega_2}
=
\frac{1}{(4\pi)^2}
\left[
1
+ \beta_1\, B_{1i} \, \hat{q}_1^i
+ \beta_2\, B_{2j} \, \hat{q}_2^j
+ \beta_1 \beta_2\,
\hat{q}_1^{\,i}\,
C_{ij}\,
\hat{q}_2^{\,j}
\right].
\end{equation}
Here, $\hat{\vec{q}}_1$ and $\hat{\vec{q}}_2$ represent the directions of the (anti)top-quark decay products used to reconstruct the top spins, while $\beta_1$ and $\beta_2$ denote the corresponding spin analysing powers. In addition, the implicit sums upon the indices $i,j$ run over the helicity-frame axes. In our analysis where only charged leptons are used, $\beta_1 = 1$ and $\beta_2 = -1$ so that Eq.~\eqref{QT} allows us to extract rather simple expressions for the components of the polarisation vectors $B$ and the spin-correlation matrix  $C$~\cite{Altakach:2022ywa},
\begin{equation}
    C_{ij} = - 9 \left\langle \cos \theta^1_i \cos \theta^2_j   \right\rangle, \qquad\qquad
    B_{1i} =   3 \left\langle \cos{\theta^1_i}   \right\rangle \qquad\qquad\text{and}\qquad\qquad
    B_{2j} = - 3 \left\langle \cos{\theta^2_j}   \right\rangle . 
\end{equation}
In these expressions, $\theta_i^{1,2}$ denote the angles between the charged lepton, evaluated in the parent (anti)top-quark rest frame, and the helicity-frame axis $i$. In practice, the elements of the spin density matrix are then extracted by filling histograms associated with the distributions of the cosine of these angles, and next by averaging the spectra over the phase space of interest. Furthermore, thanks to the SM symmetries and the choice of the helicity frame, the spin density matrix is largely diagonal in this analysis, allowing us to neglect its off-diagonal elements. This simplification preserves the dominant spin-correlation information while providing a practical and robust way to construct the leptonic angular observables that form the core of our multivariate analysis. In a second step, the availability of the spin density matrix allows for the calculation of the considered quantum-information observables, which are all functions of the elements that matrix, using the formul\ae\ defined in Section~\ref{sec:spin}. 

In addition to averages over many events, it is also possible to construct observables on an event-by-event basis using the angles determined in each individual event. Motivated by the definition of magic in Eq.~\eqref{eqn:magic}, we define the (averaged) quantity
\begin{equation}
\label{eqn:magic2p}
\widetilde{M}_2=-\log_2\, \left(\frac{
1 +\sum_{i,j\in \{k,r,n\}}\sign\left(\langle\cos{\theta_i^1} \cos{\theta_j^2}\rangle \right) \, \left(\langle \cos{\theta_i^1} \cos{\theta_j^2}\rangle \right)^4}
{ 1 +\sum_{i,j\in \{k,r,n\}}\sign\left(\langle\cos{\theta_i^1} \cos{\theta_j^2} \rangle\right) \, \left(\langle \cos{\theta_i^1} \cos{\theta_j^2}\rangle \right)^2
}\right).
\end{equation}
This variable is chosen because its sensitivity is empirically found to be slightly higher than that of the original magic $M_2$ when applied on a per-event basis instead of averaging. The corresponding event-level quantity, denoted as $\widetilde{M}^{\mathrm{evt}}_2$ for clarity, is obtained by replacing the ensemble averages appearing in Eq.~(\ref{eqn:magic2p}) by their corresponding per-event values,
\begin{equation}
\widetilde{M}_2^{\mathrm{evt}} = -\log_2\, \left(\frac{
1 +\sum_{i,j\in \{k,r,n\}}\sign\!\left(\cos{\theta_i^1} \cos{\theta_j^2}\right) \, \left(\cos{\theta_i^1} \cos{\theta_j^2}\right)^4}
{ 1 +\sum_{i,j\in \{k,r,n\}}\sign\!\left(\cos{\theta_i^1} \cos{\theta_j^2}\right) \, \left(\cos{\theta_i^1} \cos{\theta_j^2}\right)^2
}\right).
\end{equation}%
We nevertheless emphasise that neither $\widetilde{M}_2$ nor $\widetilde{M}_2^{\mathrm{evt}}$ constitutes a rigorous measure of any specific quantum-information quantity, but rather defines empirically motivated observables inspired by quantum-information studies. Similarly, the spin-correlation observables $D^{(1)}$ and $D^{(n)}$ defined in Eq.~\eqref{eq:d1} can also be evaluated on an event-by-event basis. This is achieved by expressing them in terms of the angles of the charged leptons in the parent top-quark rest frame instead of in terms of the Fano coefficients,
\begin{equation}
      D^{(1)} =- 3 \left \langle \cos{\theta^1_r} \cos{\theta^2_r}+ \cos{\theta^1_n} \cos{\theta^2_n} + \cos{\theta^1_k} \cos{\theta^2_k}  \right\rangle\qquad\text{and}\qquad
      D^{(n)} =- 3 \left\langle \cos(\theta')  \right\rangle = -3 \left \langle q^1 \cdot P
      _n(q^2)  \right \rangle,
\label{eq:Dn-costheta}
\end{equation}
where $\theta'$ is the angle between the momentum of the positive lepton and the reflected momentum of the negative lepton (with the transformation operation defined by $P_n = \mathrm{diag}(-1,-1,1)$), with both lepton momenta evaluated in the parent top-quark rest frame and the coordinates calculated in the helicity frame. We subsequently introduce the associated event-by-event variables $D^{(1)\mathrm{evt}}$ and $\cos \theta'$ (the latter being the event-level quantity whose average defines $D^{(n)}$) as the integrands of the above averages, namely
\begin{equation}
D^{(1)\mathrm{evt}} = -3 \left( \cos{\theta^1_r} \cos{\theta^2_r} + \cos{\theta^1_n} \cos{\theta^2_n} + \cos{\theta^1_k} \cos{\theta^2_k} \right)\qquad\text{and}\qquad \cos\theta' = q^1 \cdot P_n(q^2).
\end{equation}

In the next section, all of the observables introduced above, including their event-by-event implementations when relevant, are combined with various kinematic variables to assess their ability to discriminate the toponium signal from the conventional \ttbar\ background. Inputted to our BDT framework, we will show that they provide a comprehensive multivariate representation of both spin-correlation and kinematic information, enhancing the potential separation between the signal and background.

\subsection{Discriminating Signal and Background}
\label{sec:results}
\begin{figure}
  \centering
  \includegraphics[width=.5\linewidth]{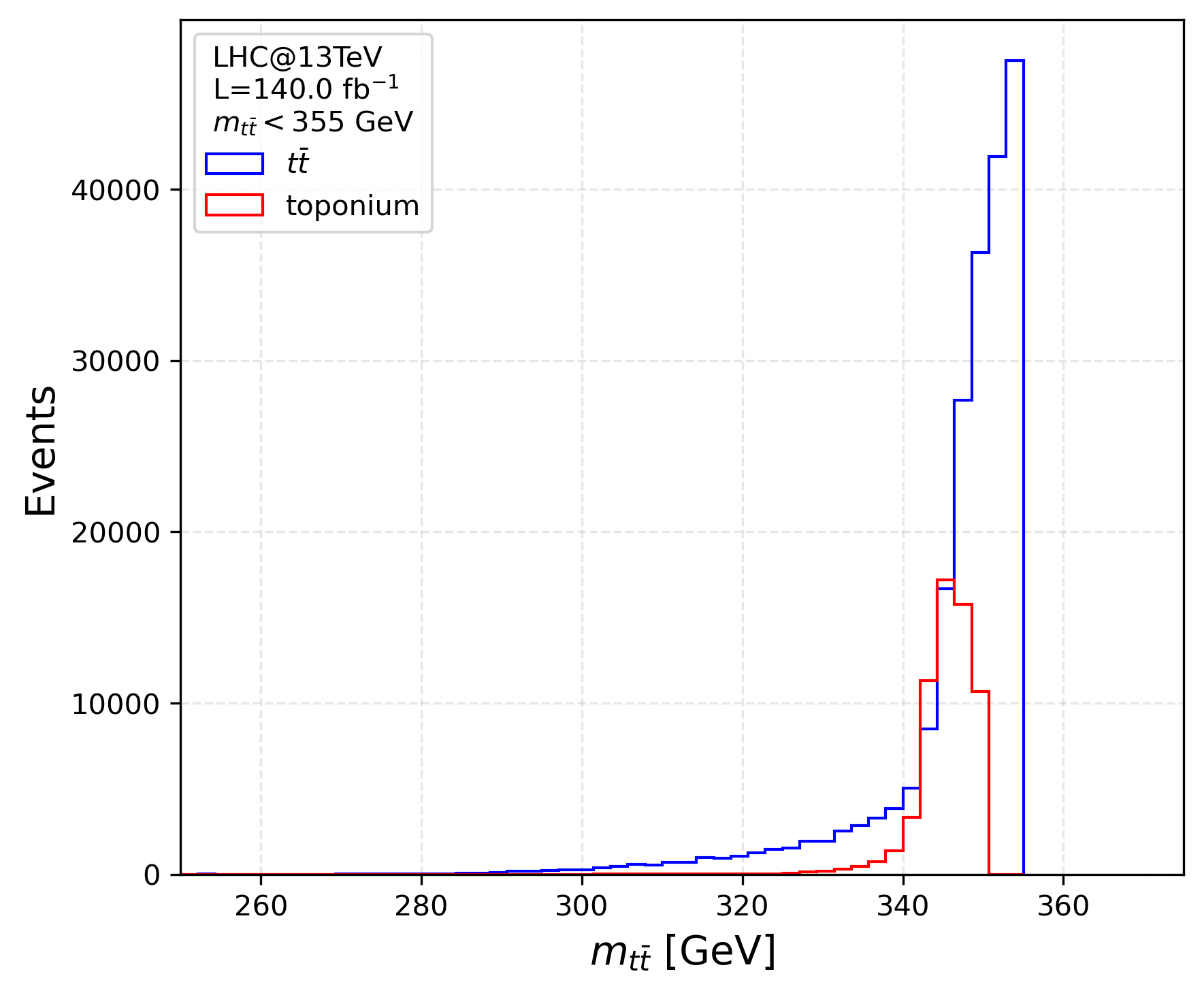}
  \caption{Distribution of the \ttbar\ invariant mass for conventional top-antitop events without any toponium contributions (blue) and toponium-only events (red). Both distributions include the fiducial cuts discussed in Section~\ref{sec:sim}, and in particular the selection $m_{t\bar t}<355$~GeV. \label{fig:mass}}
\end{figure} 

The results presented in this section are based on the simulation of LHC collisions at a centre-of-mass energy of 13~TeV. They rely on a toponium sample of 292,327 events generated in the threshold region, as well as on a continuum \ttbar\ sample from which 93,048 events satisfy the invariant-mass requirement $m_{t\bar{t}} < 355~\mathrm{GeV}$. All distributions are normalised to the expected event yields using cross sections of $\sigma_{t\bar{t}} = 833.9$~pb~\cite{Czakon:2013goa} and $\sigma_{\mathrm{toponium}} = 6.43$~pb~\cite{Sumino:2010bv, Fuks:2021xje}, an integrated luminosity of 140~fb$^{-1}$ and dileptonic top-antitop decays (with a branching fraction of approximately 0.105).  

Figure~\ref{fig:mass} shows the \ttbar\ invariant-mass distribution for the toponium sample (red) and the conventional continuum \ttbar\ background (red) after applying the selection described in Section~\ref{sec:sim}. As expected, the toponium distribution is strongly concentrated near the threshold region, reflecting the quasi-bound-state nature of the top-antitop pair and the low relative velocity required for binding. In contrast, the continuum \ttbar\ events exhibit a broader invariant-mass spectrum that falls smoothly near threshold due to  phase-space suppression. While applying the invariant-mass cut $m_{t\bar t} < 355~\mathrm{GeV}$ enhances the relative contribution of the toponium signal within the selected region, the absolute event yield remains small compared to the continuum. This observation underscores the importance of employing additional observables, such as spin correlations and quantum-information-inspired variables, to further distinguish toponium events from the conventional background.

\begin{table}
  \renewcommand{\arraystretch}{1.4}\setlength{\tabcolsep}{12pt}
  \centering
\begin{tabular}{
  l
  S[table-format=-2.4(4)]
  S[table-format=-2.4(4)]
  S[table-format=-2.4(4)]
}
\toprule
& {$t\bar t$} 	& {Toponium} & {Combined} \\
\midrule 
$C_{kk}$        & -0.5003 +- 0.0097 & -0.9594 +- 0.0053 & -0.5954 +- 0.0077 \\ 
$C_{kr}$        &  0.0635 +- 0.0098 &  0.0113 +- 0.0056 &  0.0527 +- 0.0079 \\ 
$C_{kn}$        & -0.0034 +- 0.0099 &  0.0027 +- 0.0055 & -0.0021 +- 0.0079 \\ 
$C_{rk}$        &  0.0766 +- 0.0098 &  0.0094 +- 0.0056 &  0.0626 +- 0.0079 \\ 
$C_{rr}$        & -0.3903 +- 0.0098 & -0.8965 +- 0.0053 & -0.4951 +- 0.0078 \\ 
$C_{rn}$        & -0.0145 +- 0.0098 & -0.0040 +- 0.0055 & -0.0123 +- 0.0079 \\ 
$C_{nk}$        &  0.0133 +- 0.0098 & -0.0030 +- 0.0056 &  0.0100 +- 0.0078 \\ 
$C_{nr}$        &  0.0007 +- 0.0098 &  0.0030 +- 0.0055 &  0.0012 +- 0.0079 \\ 
$C_{nn}$        & -0.5547 +- 0.0097 & -0.9224 +- 0.0053 & -0.6308 +- 0.0077 \\ 
\midrule 
$D^{(1)}$       & -0.4818 +- 0.0057 & -0.9261 +- 0.0031 & -0.8188 +- 0.0027 \\ 
$D^{(k)}$       &  0.1482 +- 0.0056 &  0.2865 +- 0.0030 &  0.2531 +- 0.0027 \\ 
$D^{(n)}$       &  0.2216 +- 0.0056 &  0.3284 +- 0.0030 &  0.3026 +- 0.0027 \\ 
$D^{(r)}$       &  0.1120 +- 0.0056 &  0.3112 +- 0.0031 &  0.2631 +- 0.0027 \\ 
$\mathcal{C}$   &  0.2234 +- 0.0086 &  0.8930 +- 0.0050 &  0.7293 +- 0.0043 \\ 
${M}_2$   		&  0.5481 +- 0.0051 &  0.1525 +- 0.0069 &  0.3537 +- 0.0047 \\ 
$\widetilde{M}_2$ & 0.0124 +- 0.0003 & 0.0447 +- 0.0003 & 0.0351 +- 0.0002 \\ 
$\mu(\rho)$		&  0.2421 +- 0.0056 &  0.8586 +- 0.0062 &  0.6723 +- 0.0044 \\ 
$E_{N}(\rho)$   &  0.2906 +- 0.0100 &  0.9178 +- 0.0038 &  0.7894 +- 0.0034 \\ 
$D_{T}(\rho)$   &  0.0000 +- 0.0000 &  0.3338 +- 0.0050 &  0.2543 +- 0.0047 \\ 
  \bottomrule
  \end{tabular}
  \caption{Spin-correlation coefficients $C_{ij}$ and quantum-information-inspired observables for dileptonic $t\bar t$ production in the threshold region. Results are shown for the SM continuum \ttbar\ sample, the toponium-only sample, and their weighted combination according to the corresponding cross sections. \label{tab:qimatrix}}
\end{table}

Table~\ref{tab:qimatrix} summarises our results for the elements of the spin-correlation matrix and the quantum-information-inspired observables for the SM \ttbar\ continuum, the toponium sample and their weighted combination according to the cross sections given above. Due to parity and CP invariance, the components of the single-spin polarisation coefficients $B_1$ and $B_2$ vanish when integrated over the full phase space. In our Monte Carlo simulations, the event-averaged $B$ vectors acquire, however, small but non-zero values. These arise primarily from finite event statistics, fiducial phase-space cuts and the use of event-dependent spin-quantisation axes, which collectively prevent exact cancellation. More precisely, the magnitude of the vector components are found more than one order of magnitude smaller than the diagonal elements of the spin-density matrix, so they are not considered further in this analysis.

The diagonal elements of the spin-correlation matrix $C_{ii}$ (with $i=n, k, r$) show strikingly different behaviour between the toponium and continuum samples. In the toponium sample, the $C$ matrix is close to $-\mathbb{I}_3$, reflecting the formation of a coherent, quasi-bound top-antitop system near threshold with well-defined total quantum numbers. In this regime, the single-spin degrees of freedom are suppressed, as above-mentioned, while the relative spin correlations are strongly enhanced. By contrast, continuum \ttbar\ events correspond to more independent top and antitop quarks, where spin correlations are comparatively weaker. This is reflected, for instance, in the trace distance $D_T(\rho)$ between the spin correlation matrices, which is substantial. The enhanced spin correlations in the toponium sample propagate directly to all derived quantum-information observables. The coefficients $D^{(i)}$, which depend on the diagonal components of the spin-correlation matrix, are significantly larger than in the continuum case, as is the case for concurrence $\mathcal{C}$, the normalised purity $\mu(\rho)$, the logarithmic negativity $E_N(\rho)$ and the inspired-magic number $\widetilde{M}_2$. All of these observables thus exhibit a clear separation power between the \ttbar-only scenario and the combination including toponium, highlighting their potential for enhancing discrimination of near-threshold top-antitop bound-state effects.

To further characterise the geometric relationship between the reconstructed density matrices $\rho_{\rm sm}$ and $\rho_{\rm toponium}$, we evaluate the Hilbert-Schmidt cosine between their deviations from the maximally mixed state, $\Delta\rho \equiv \rho - \mathbb{I}_4/4$,
	\begin{equation}
		\cos\alpha = \frac{{\rm Tr}\!\left(\Delta\rho_{\mathrm{sm}}\,\Delta\rho_{\rm toponium}\right)}{\sqrt{{\rm Tr}\!\left(\Delta\rho_{\mathrm{sm}}^2\right)\,{\rm Tr}\!\left(\Delta\rho_{\rm toponium}^2\right)}} = 0.9826,
	\end{equation}
which is very close to unity. This indicates that the SM and toponium density matrices are nearly collinear in operator space relative to the maximally mixed state. In other words, the primary difference between the two samples arises mainly from the magnitude of their deviation from $\mathbb{I}_4/4$, rather than a change in the direction of that deviation. This near-collinearity provides a natural geometric interpretation of the trace distance $D_{T}(\rho)$ reported in Table~\ref{tab:qimatrix}, showing that the discrimination power is largely driven by the overall distance from the maximally mixed state which encodes the strength of spin correlations.

\begin{figure}
\centering
  \includegraphics[width=.48\linewidth]{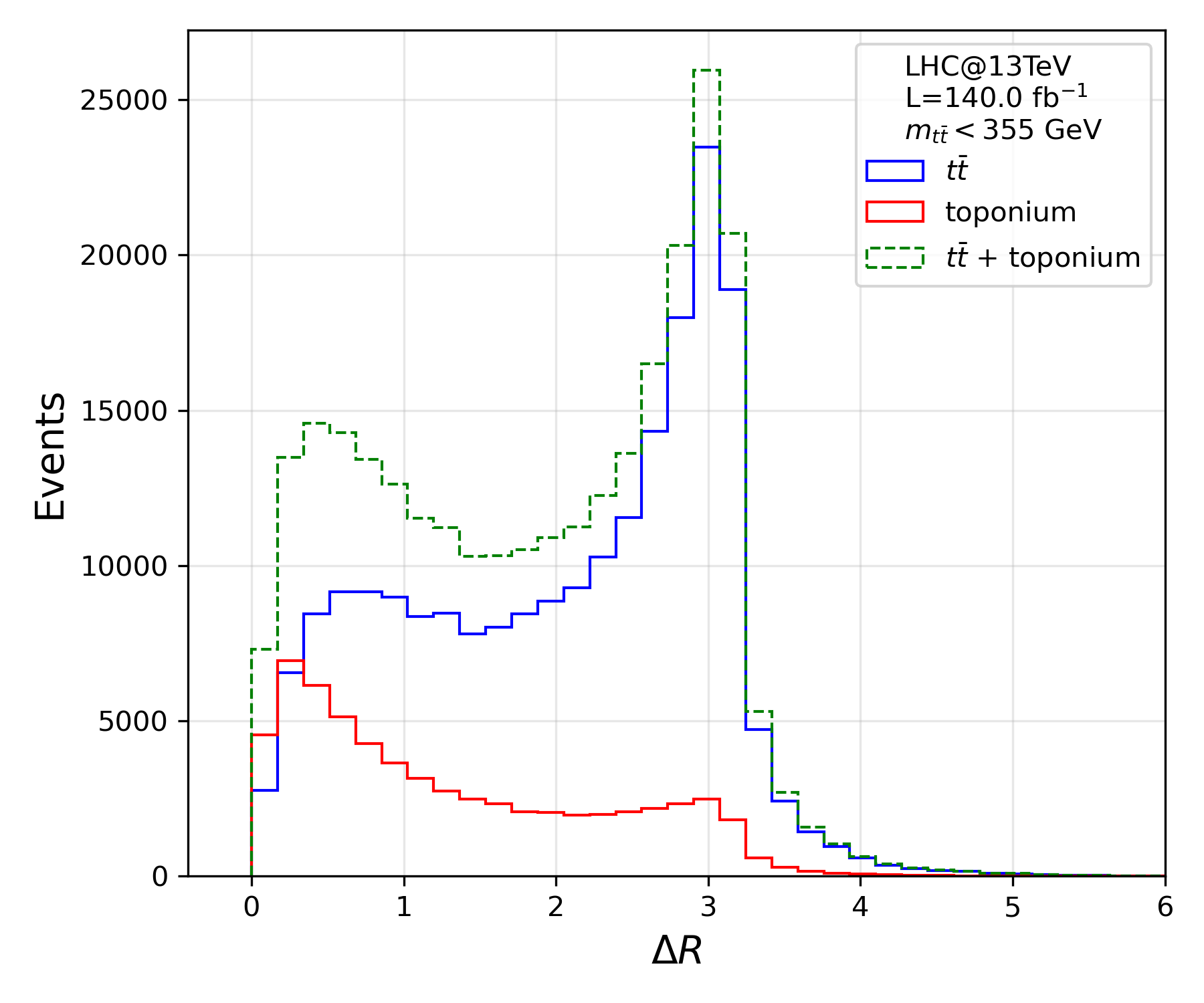}\hfill
  \includegraphics[width=.48\linewidth]{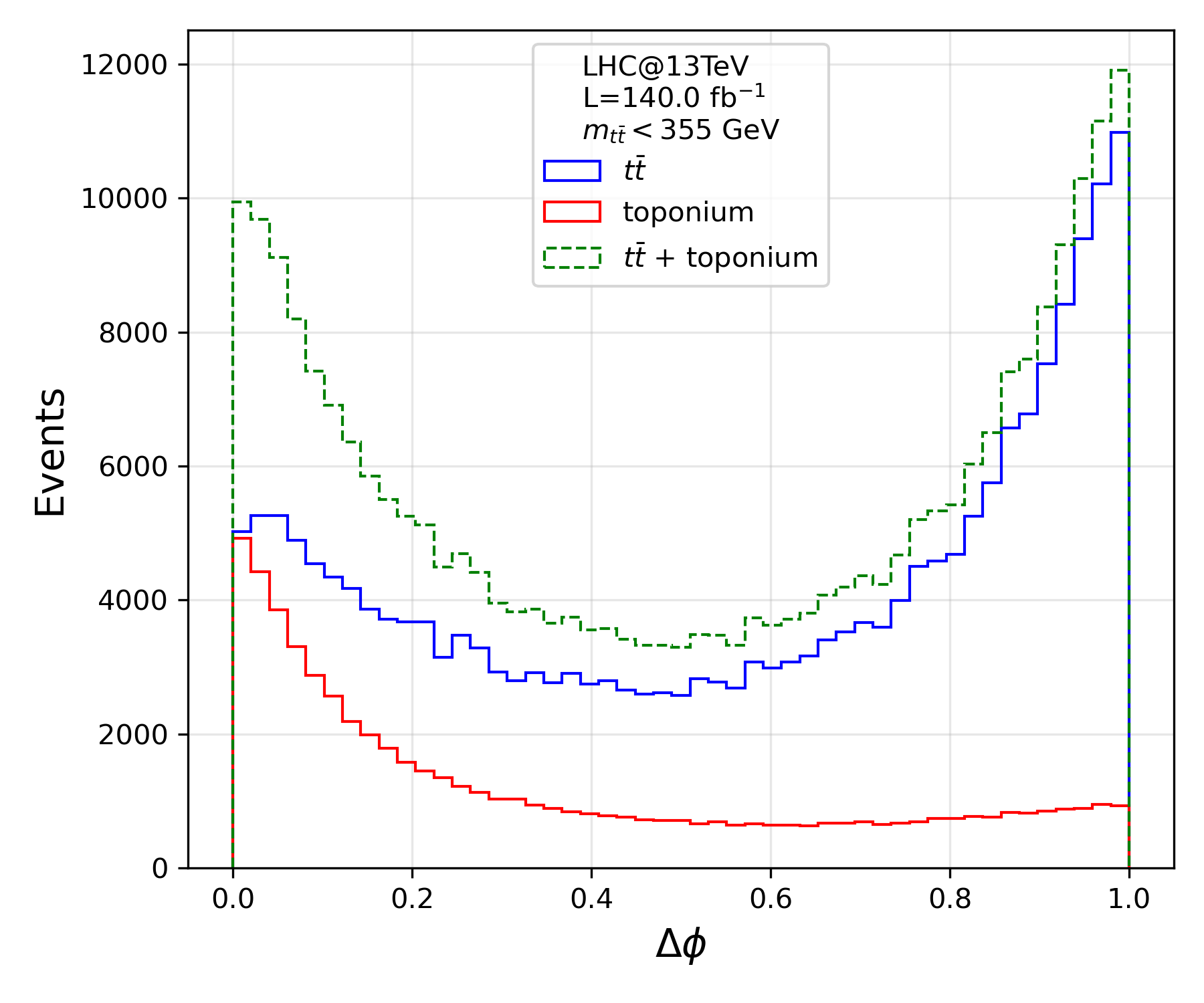}\\
  \includegraphics[width=.48\linewidth]{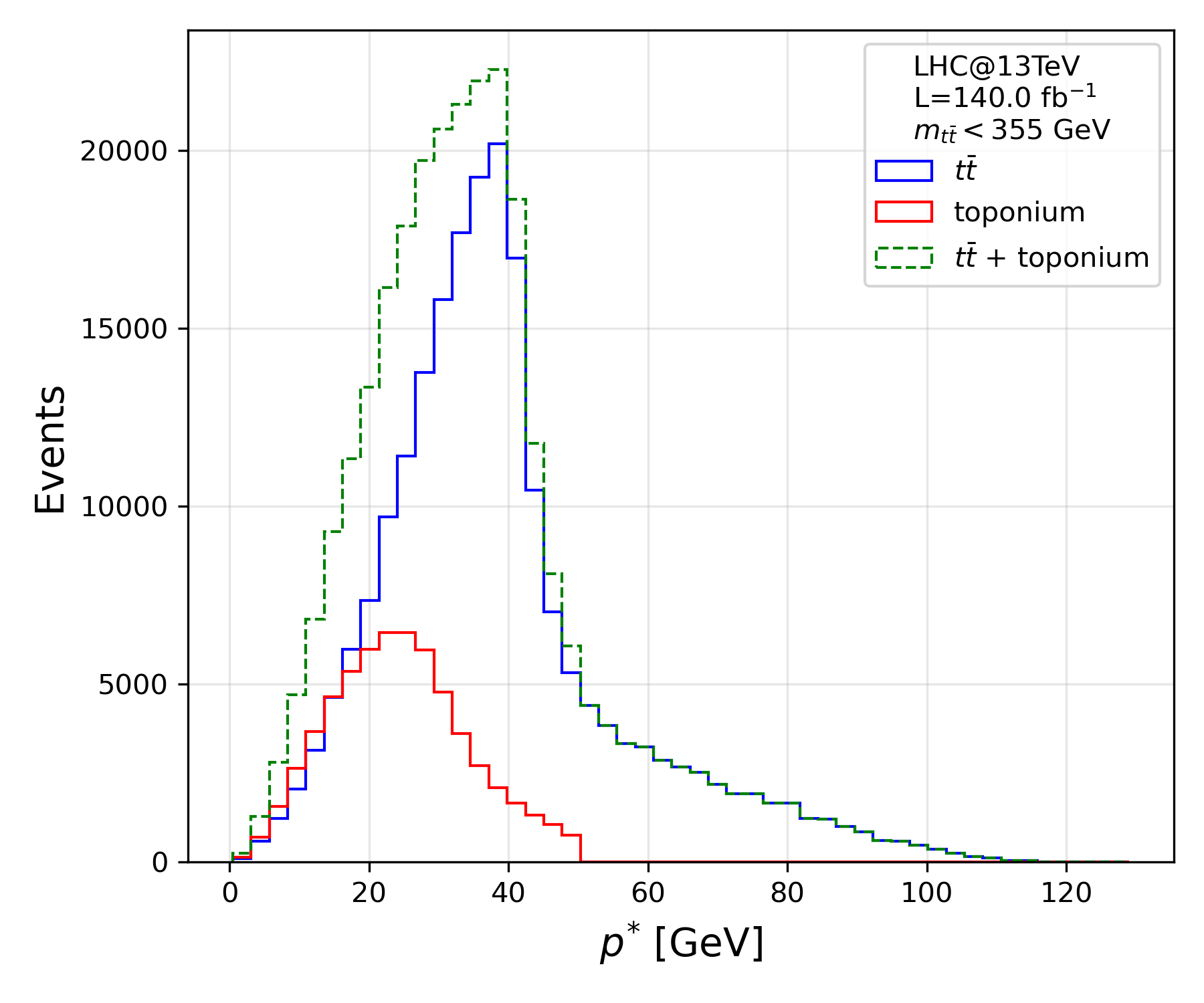}\hfill
  \includegraphics[width=.48\linewidth]{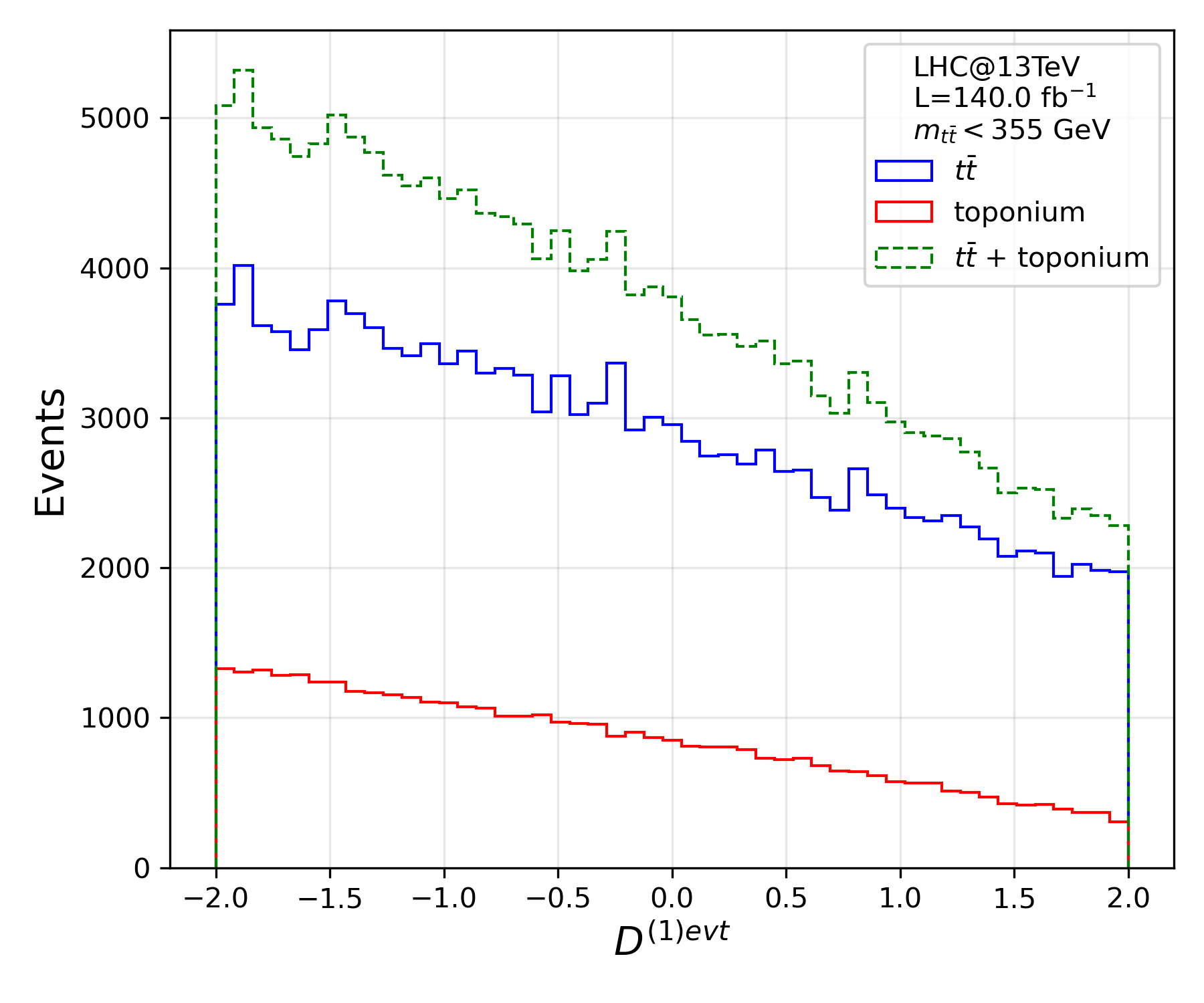}\\
  \includegraphics[width=.48\linewidth]{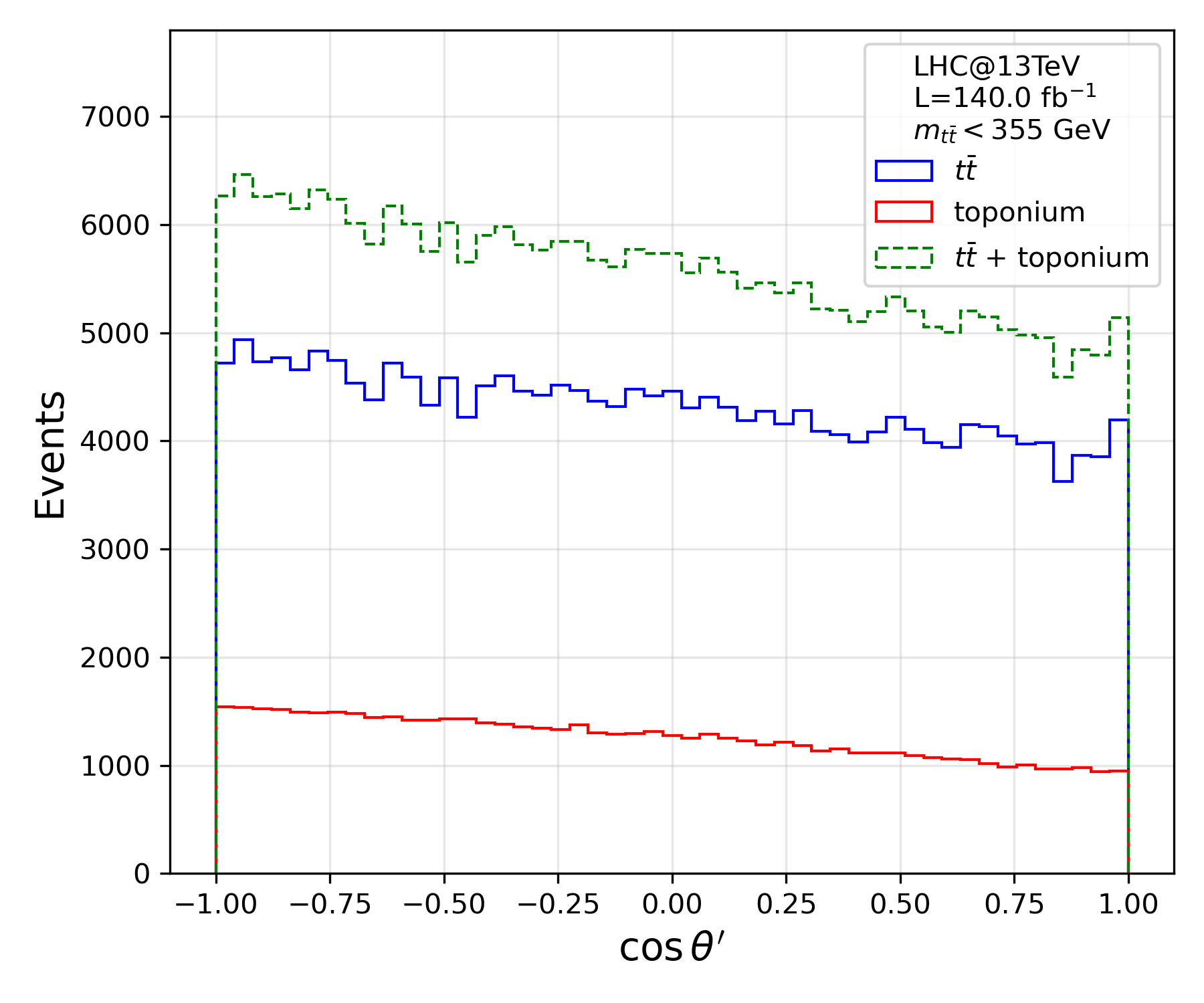}\hfill
  \includegraphics[width=.48\linewidth]{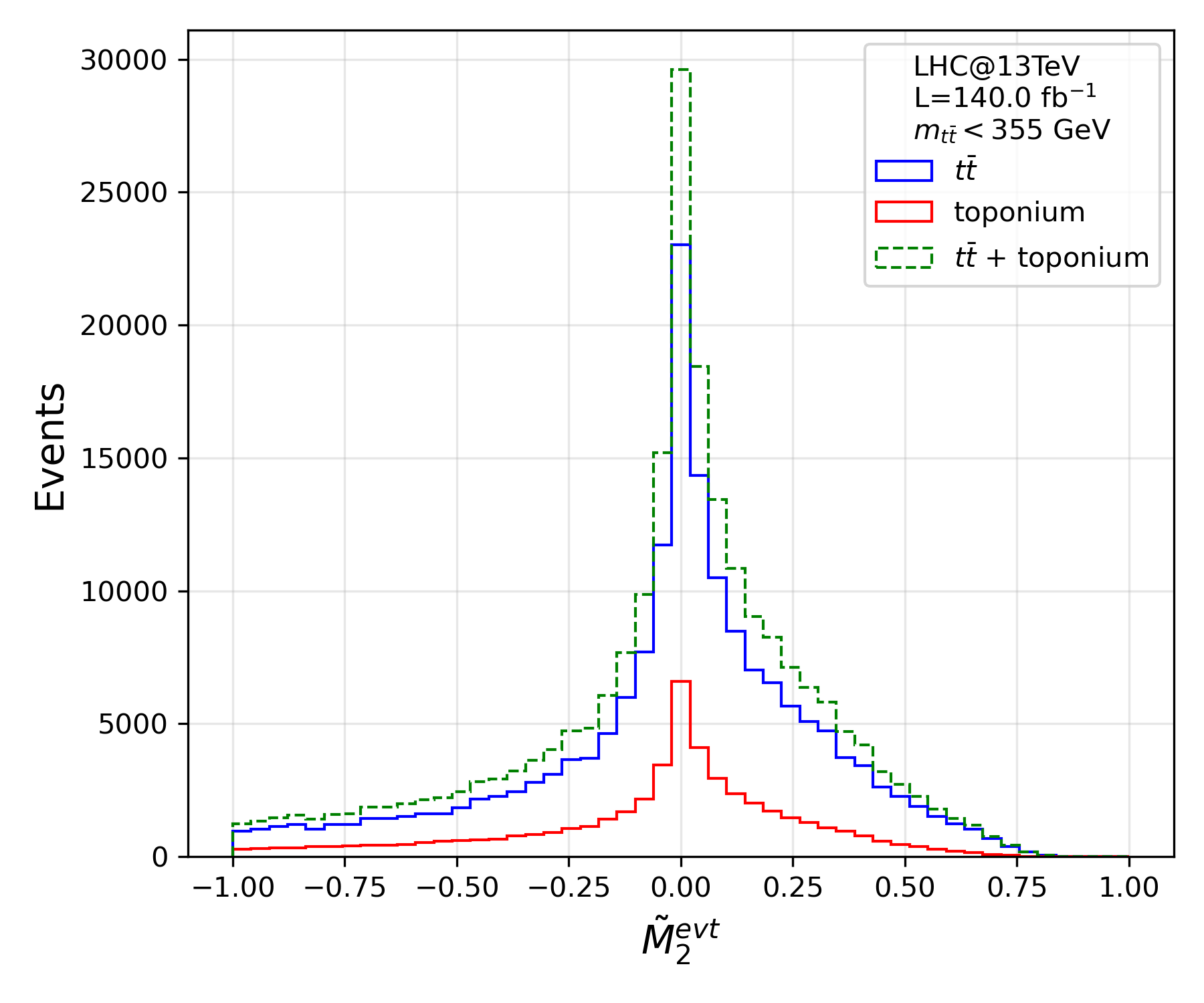}
  \caption{Various distributions for conventional \ttbar\ events without toponium effects (solid blue), toponium-only events (solid red) and \ttbar\ events combined with the toponium contributions (dashed green). We present results for the $\Delta R$ (top left) and $\Delta\phi$ (top right) angular separation between the two final-state leptons, the $p^\star$ (middle left), $D^{(1)\mathrm{evt}}$ (middle right), $\cos \theta'$ (bottom left) and $\widetilde{M}^{\mathrm{evt}}_2$ (bottom right) observables. The results are normalised to $\mathcal{L}=140$~fb$^{-1}$ of LHC collisions at 13~TeV and include the selection cuts of Section~\ref{sec:sim}. \label{fig:kin_QI}}
\end{figure}

\begin{figure}
  \centering
  \includegraphics[width=.48\linewidth]{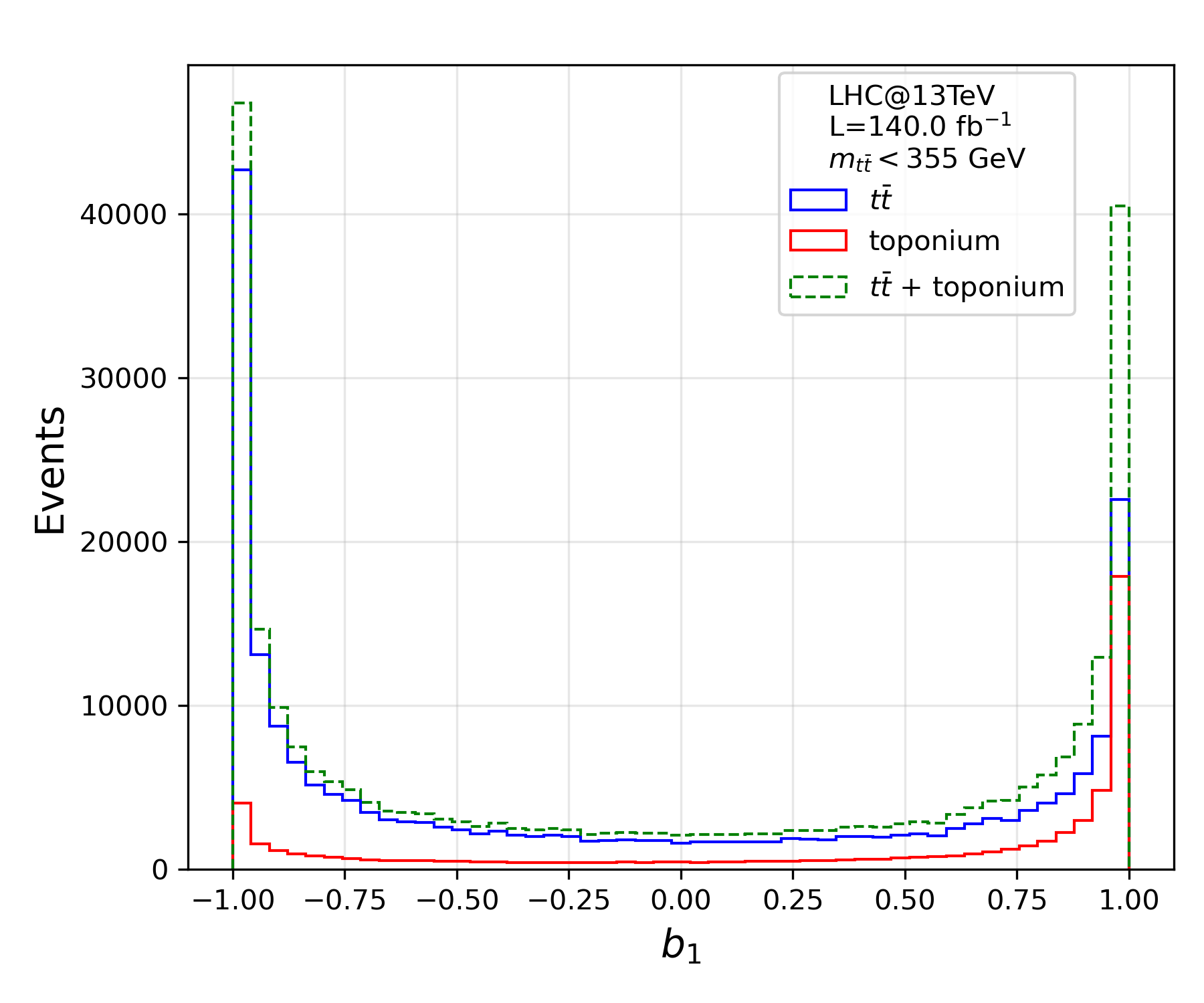}\hfill
  \includegraphics[width=.48\linewidth]{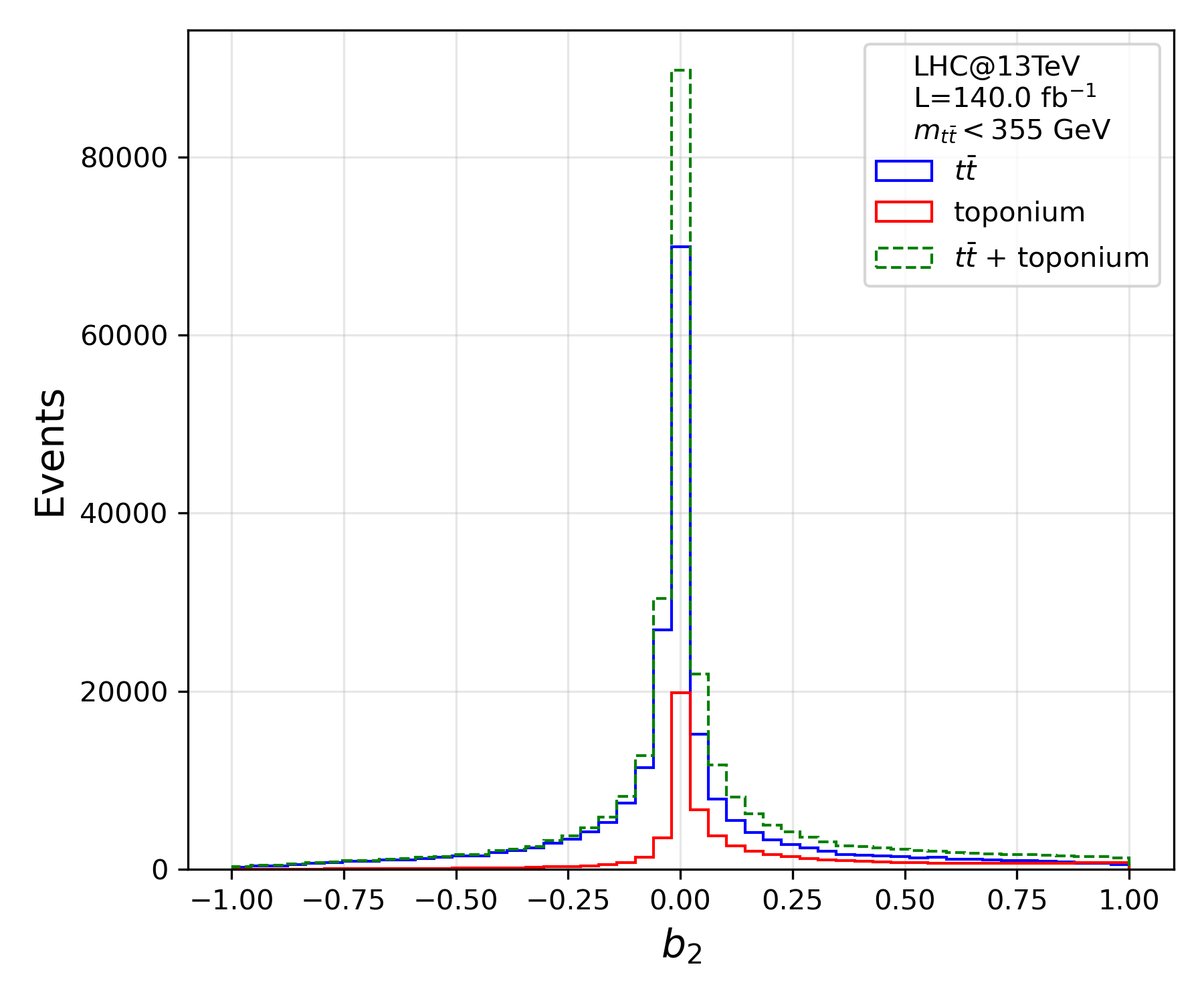}\\
  \includegraphics[width=.48\linewidth]{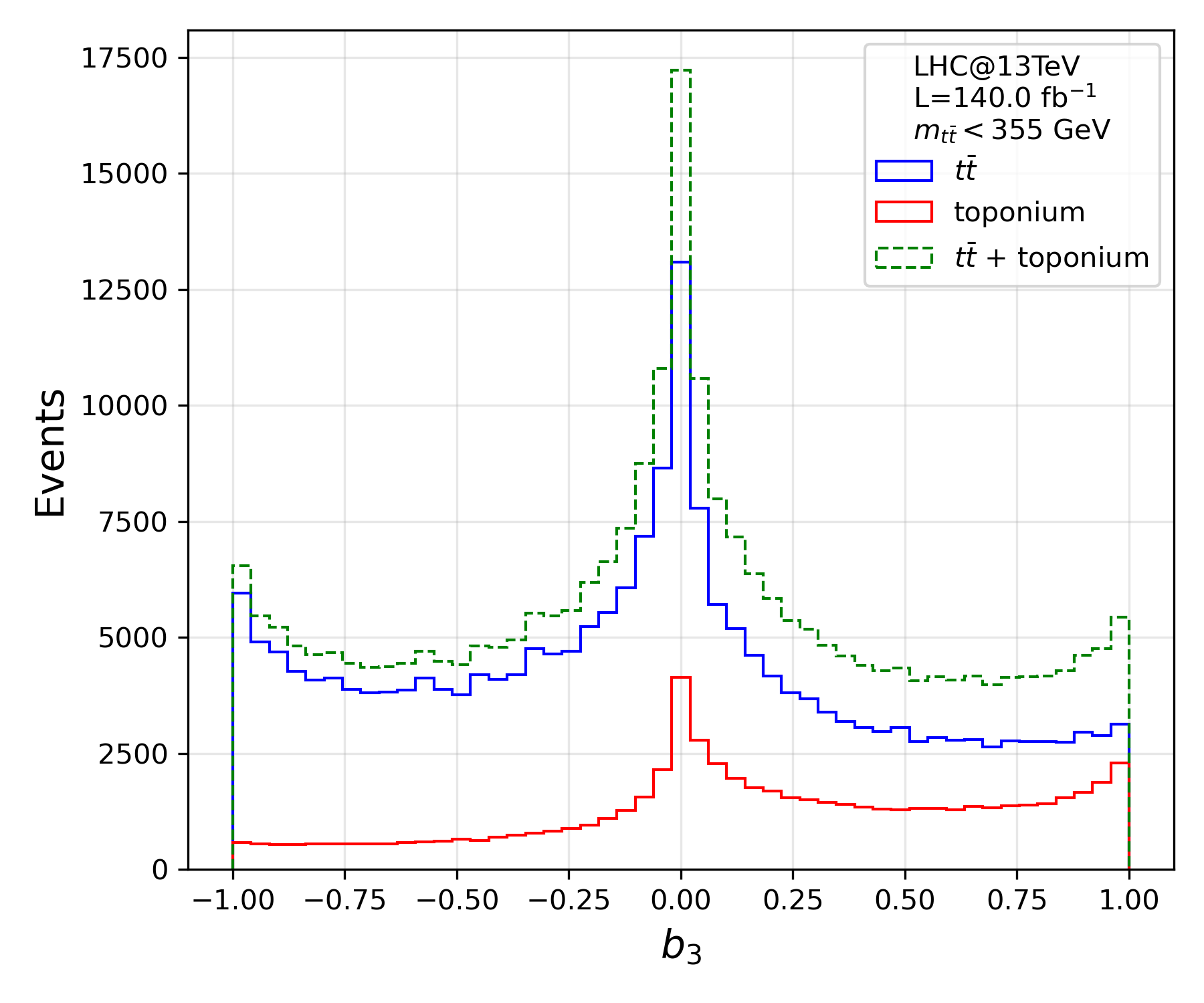}\hfill
  \includegraphics[width=.48\linewidth]{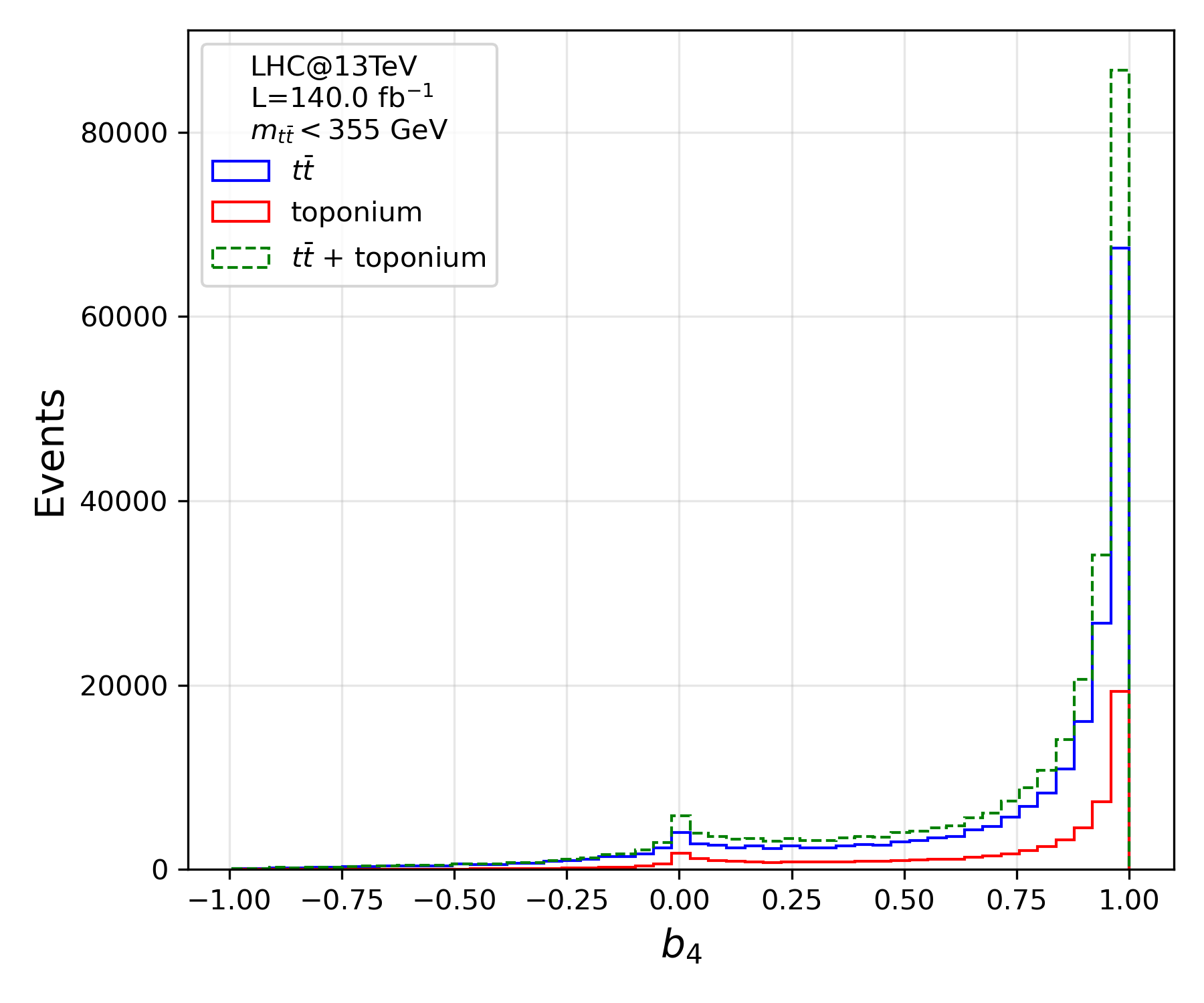}
  \caption{Distributions of the variables $b_1$, $b_2$, $b_3$ and $b_4$ for conventional \ttbar\ events without toponium effects (solid blue), toponium-only events (solid red) and \ttbar\ events combined with the toponium contributions (dashed green). The results are normalised to $\mathcal{L}=140$~fb$^{-1}$ of LHC collisions at 13~TeV and include the selection cuts of Section~\ref{sec:sim}. \label{fig:b}}
\end{figure}

Figures~\ref{fig:kin_QI} and \ref{fig:b} show the normalised distributions of the main observables sensitive to toponium effects, evaluated after applying the analysis selection discussed in Section~\ref{sec:sim}. We present predictions for the conventional $t\bar t$ continuum (solid blue), the toponium-only event sample (solid red) and their combination (dashed green). The differences between the samples are strongly highlighted in the angular and kinematic observables of Figure~\ref{fig:kin_QI}. Toponium events typically produce top-quark pairs that are closer in phase space, yielding more collimated leptons with lower distance in the transverse plane $\Delta R$ and smaller azimuthal angle separation $\Delta\phi$. The magnitude of the momentum of the top in the \ttbar\ rest frame $p^\star$ is correspondingly smaller for toponium events, consistent with the slow relative motion necessary for binding with a distribution featuring a peak close to the inverse of the associated Bohr radius. In addition, quantum-information-inspired observables such as $D^{(1)\mathrm{evt}}$, $\cos\theta'$ and $\widetilde{M}^{\mathrm{evt}}_2$ show moderate but noticeable shifts relative to the continuum. These observables encode the spin correlations of the top-quark pair, which are enhanced in toponium formation due to the presence of a coherent two-particle state. The variables $b_1$, $b_2$, $b_3$ and $b_4$, displayed in Figure~\ref{fig:b}, further exhibit the differences among the samples. In particular, $b_1$ and $b_4$ tend to peak near unity (in absolute value), while $b_2$ and $b_3$ accumulate near zero for the toponium sample. Since these variables are constructed purely from the top-quark momenta, this behaviour reflects the enhanced angular correlations between the top and antitop in the quasi-bound toponium system, where the relative motion is constrained near threshold. The continuum \ttbar events, by contrast, display broader distributions, consistent with the more independent production of top quarks in the high-energy scattering regime captured by the wide selection cut $m_{t\bar t}<355$~GeV. 

\begin{figure}
  \centering
  \includegraphics[width=0.46\linewidth]{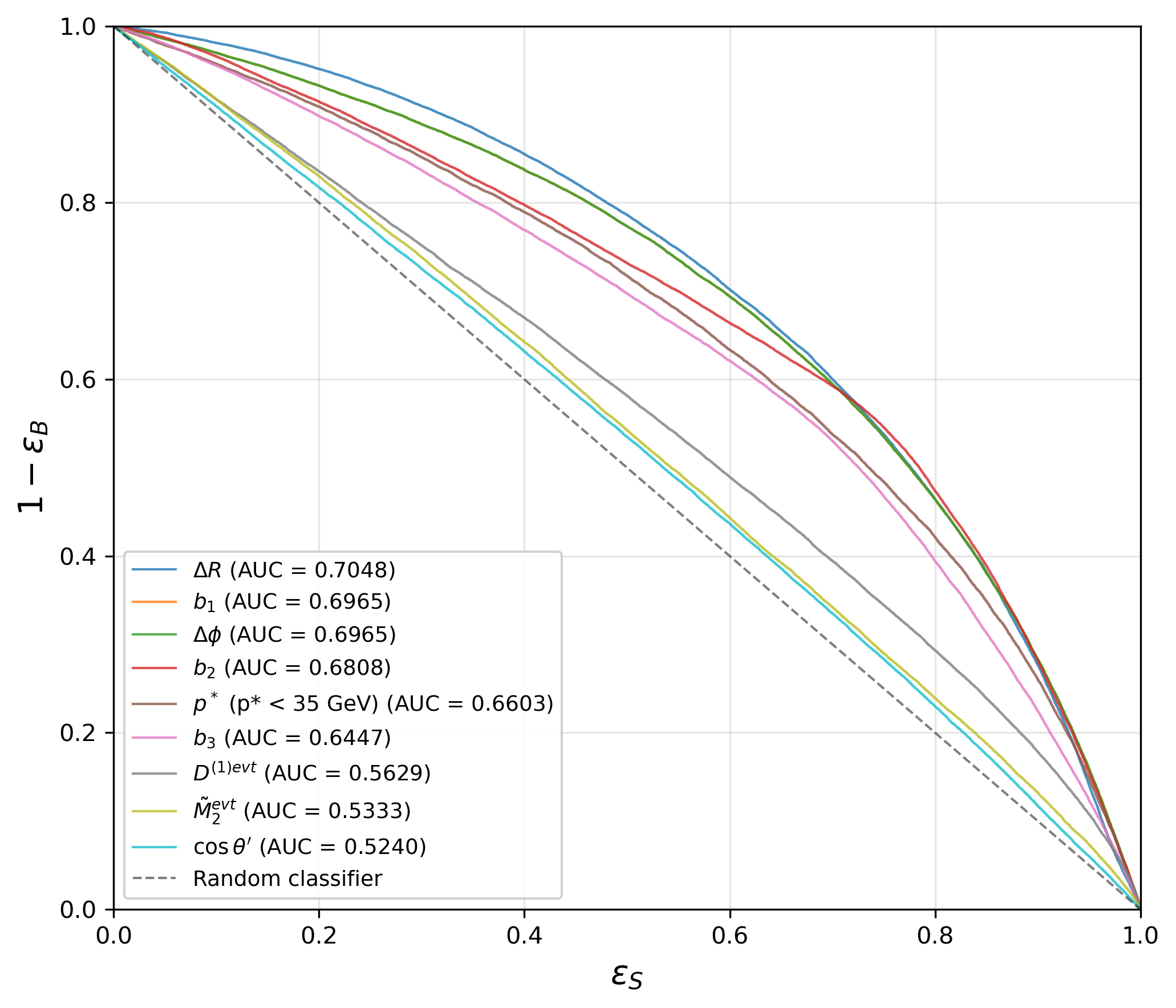}\hfill
  \includegraphics[width=0.48\linewidth]{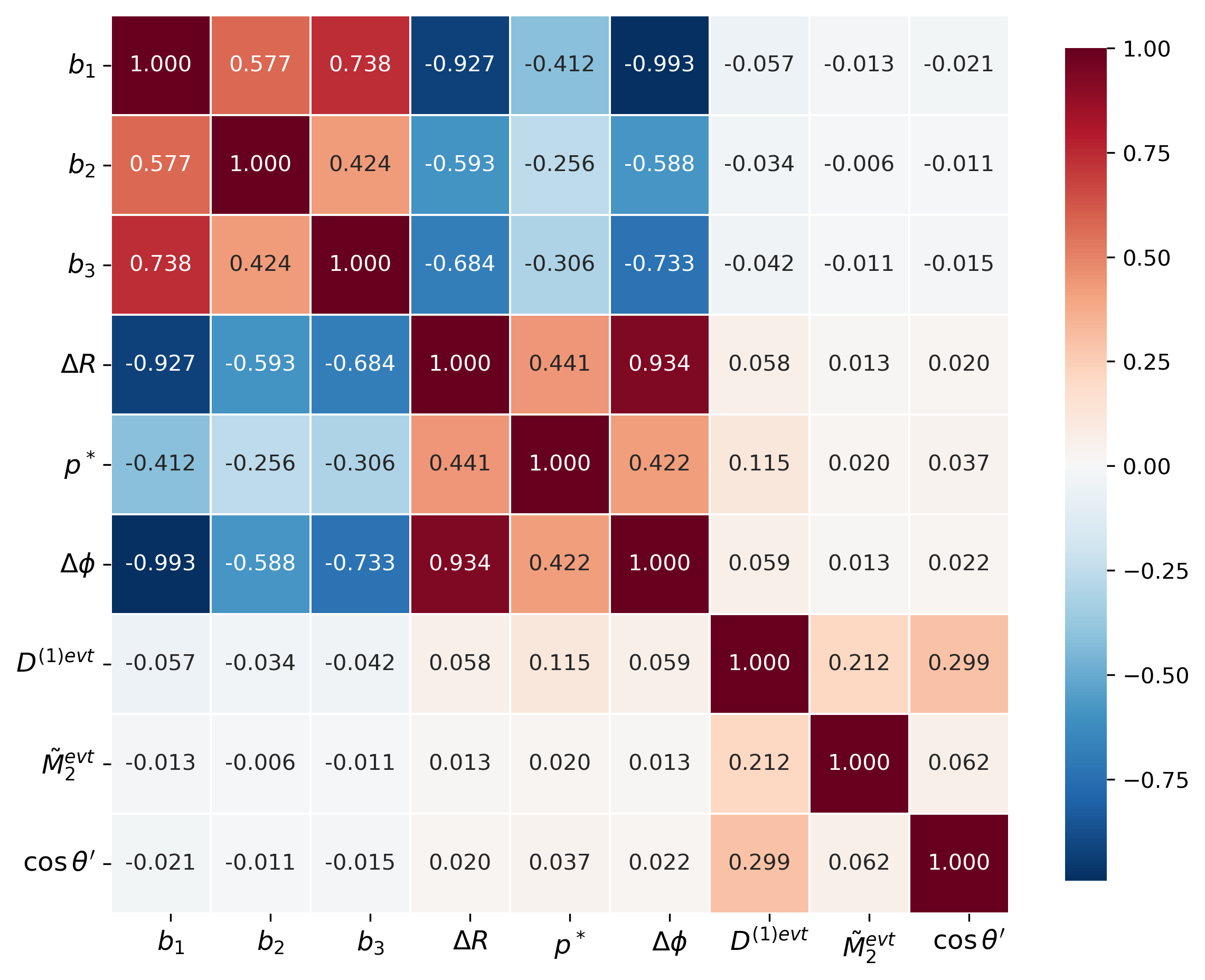}\\[.3cm]
  \includegraphics[width=0.99\linewidth]{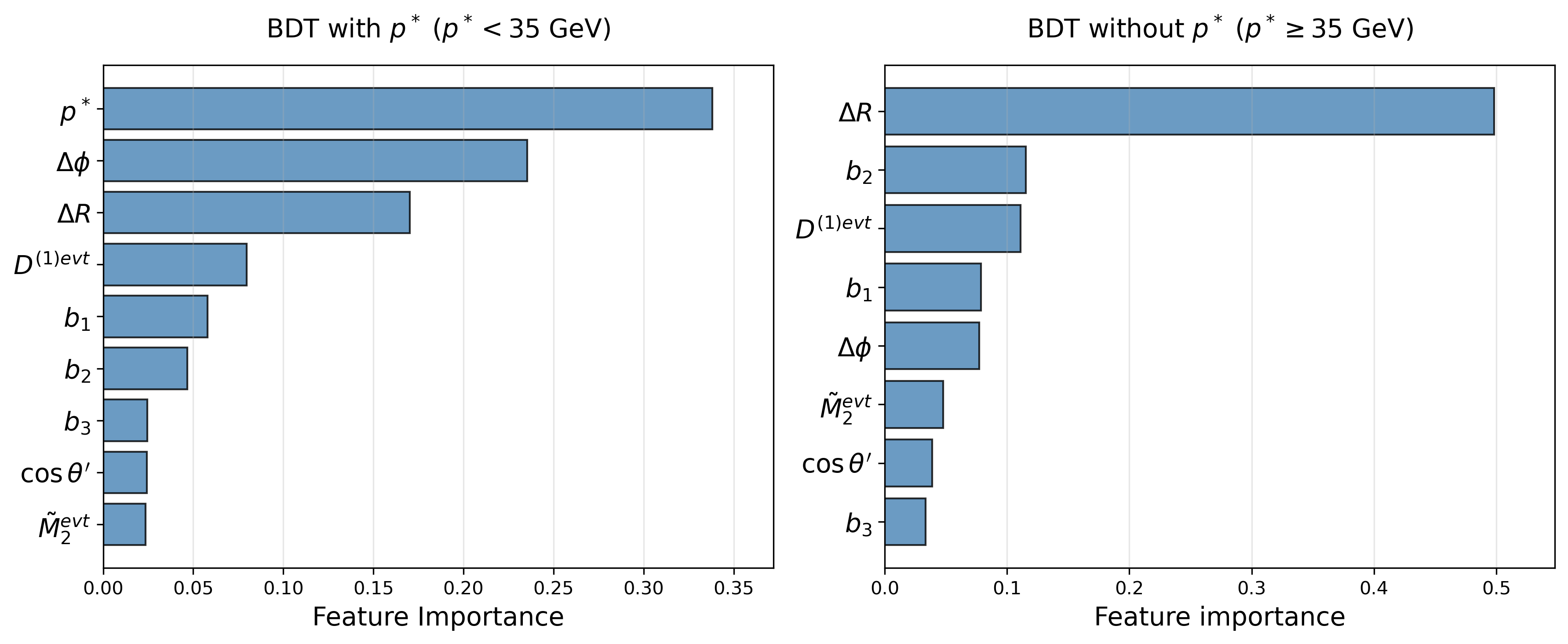}
  \caption{Performance and correlations of the input variables used in our BDT analysis of the $t\bar t$ threshold region. \textit{Top left} -- ROC curves for the individual observables, showing the \ttbar\ background rejection $(1-\varepsilon_B)$ as a function of the toponium signal efficiency $(\varepsilon_S)$; the corresponding AUC values are given in the legend. \textit{Top right} -- Correlation matrix of all BDT input variables after event selection. \textit{Bottom panels} -- Feature importance obtained from the two BDT strategies. The bottom-left panel corresponds to events with $p^\star < 35$~GeV, where $p^\star$ is included among the inputs, while the bottom-right panel corresponds to events with $p^\star \geq 35$~GeV, where $p^\star$ is excluded. \label{fig:BDT_summary}}
\end{figure} 

Taken together, the kinematic and spin-correlation observables provide complementary information: the kinematic variables capture the constrained phase-space configuration of near-threshold production, while the quantum-information-inspired observables probe the enhanced spin correlations of the bound state. This combination motivates their use as inputs to the BDT classifiers built in this work, which exploit both aspects to maximise the separation between the toponium signal and the conventional \ttbar\ background. We begin by assessing the discriminating power of each individual observable in the top left panel of Figure~\ref{fig:BDT_summary}, which shows the Receiver Operating Characteristic (ROC) curves for all input variables considered. The curves are displayed as a function of the toponium signal efficiency $\varepsilon_S$ and the corresponding \ttbar\ continuum rejection $(1-\varepsilon_B)$, and the associated Area Under the Curve (AUC) provides a quantitative measure of the separation performance. Among the considered observables, the dileptonic angular variables $\Delta R$ and $\Delta\phi$ describing the separation between the two final-state leptons in the transverse plane, together with $b_1$, yield the strongest single-variable discrimination. This behaviour is consistent with the near-threshold kinematics of toponium production where the top quarks are produced with small relative momentum and therefore exhibit characteristic angular configurations. Furthermore, the ROC curves for the $b_1$ and $\Delta\phi$ observables are identical, reflecting their very strong correlation as confirmed by the correlation matrix given in the right panel of the figure. The observables $b_2$, $b_3$ and $p^\star$ also exhibit non-negligible discriminating power, indicating that variables constructed purely from the top-quark momenta retain a potential sensitivity to toponium formation. In contrast, the quantum-information-inspired observables $D^{(1)\mathrm{evt}}$, $\cos\theta'$ and $\widetilde{M}^{\mathrm{evt}}_2$ show limited separation when used individually, with AUC values close to the random-classifier expectation. Finally, it turns out that the variable $b_4$ does not exhibit significant discriminating power. It is therefore excluded from the analysis. This hierarchy indicates that global angular and kinematic observables provide the strongest discrimination at the single-variable level, and therefore motivates the use of a multivariate approach to exploit the existing correlations among the variables and combine any complementary information. 

The correlation matrix of the BDT inputs is shown in the top right panel of Figure~\ref{fig:BDT_summary}. It demonstrates that the considered observables cluster into well-defined groups. The variables $b_1$, $b_2$ and $b_3$ are strongly correlated with each other and show pronounced (anti-)correlations with the angular observables $\Delta R$ and $\Delta\phi$. As expected, $\Delta R$ and $\Delta\phi$ are themselves strongly correlated, indicating that they probe closely related features of the dileptonic angular configuration. By contrast, the quantum-information-inspired observables $D^{(1)\mathrm{evt}}$, $\cos\theta'$ and $\widetilde{M}^{\mathrm{evt}}_2$ exhibit only weak correlations with the dominant angular variables, while showing moderate correlations among themselves. This pattern suggests that these observables encode information that is not fully captured by the purely kinematic variables. The coexistence of strong intra-group correlations and weaker inter-group correlations once again supports our multivariate approach to exploit both the dominant kinematic differences and the subleading spin-correlation effects.

The feature ranking obtained from the two BDT strategies is shown in the bottom row of Figure~\ref{fig:BDT_summary}. It is evaluated using the average gain metric provided by the \textsc{XGBoost} framework, which quantifies the average reduction in the loss function (log-loss) determined when a feature is used to perform a split in the decision trees. For each feature, the gain is accumulated over all splits in which it appears and averaged over those splits, and the resulting values are normalised to unity for presentation. For events in which the magnitude of the top momentum in the $t\bar t$ rest frame $p^\star < 35$~GeV, that corresponds to the case where $p^\star$ is included as an input to the BDT analysis, it is ranked as the most important variable followed by $\Delta R$ and $\Delta\phi$. This reflects the strong sensitivity of $p^\star$ to near-threshold production. Interestingly, $D^{(1)\mathrm{evt}}$ appears above $b_1$, $b_2$ and $b_3$ in the ranking, demonstrating that spin-sensitive observables contribute non-negligibly once correlations among variables are taken into account. In the complementary region in which $p^\star \geq 35$~GeV, where $p^\star$ is excluded from the BDT inputs, the angular observables dominate the ranking while the quantum-information-inspired variables retain a visible, though subleading, contribution.

\begin{figure}
  \centering
   \includegraphics[width=0.48\linewidth]{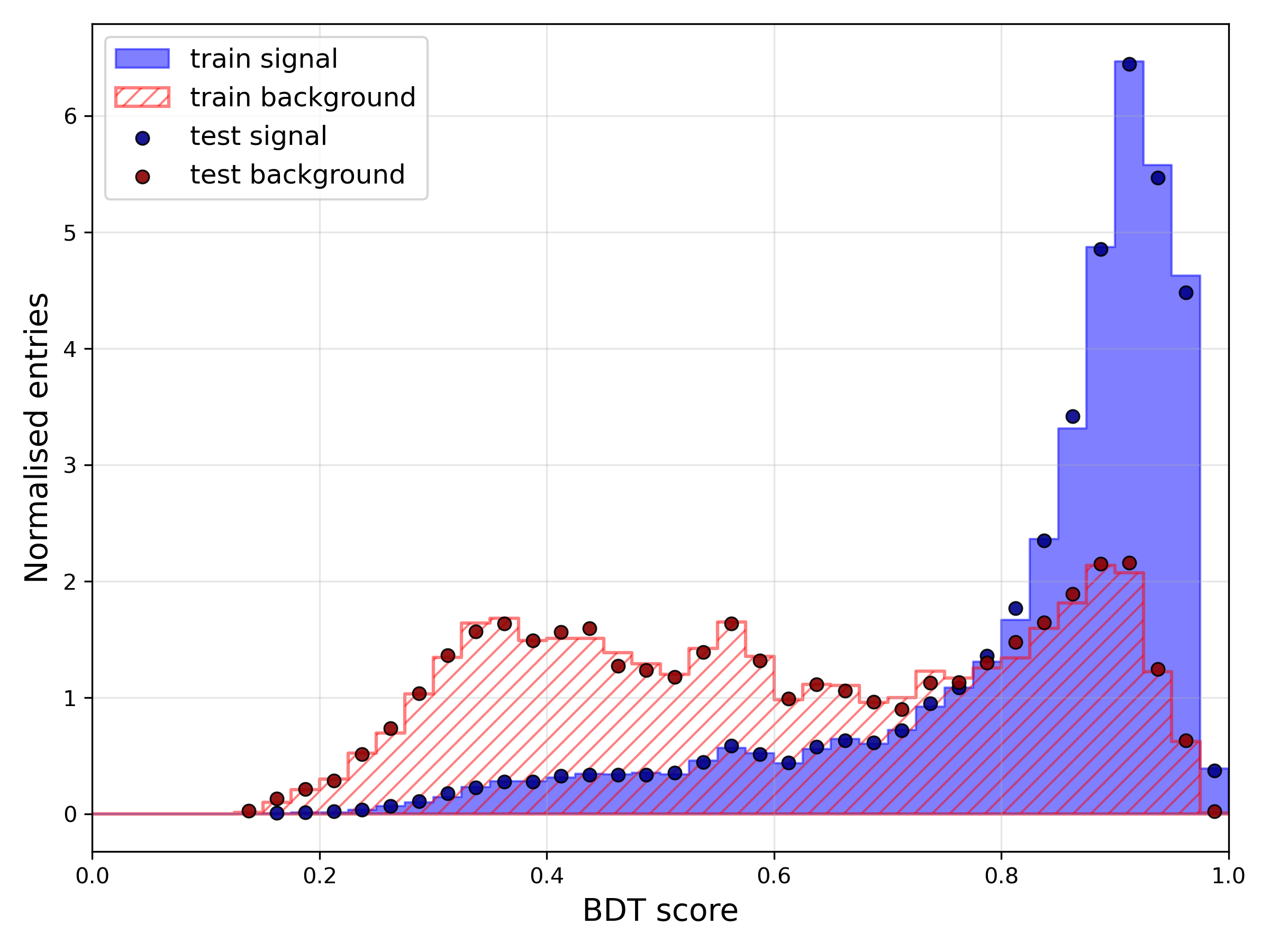} \hfill
   \includegraphics[width=0.48\linewidth]{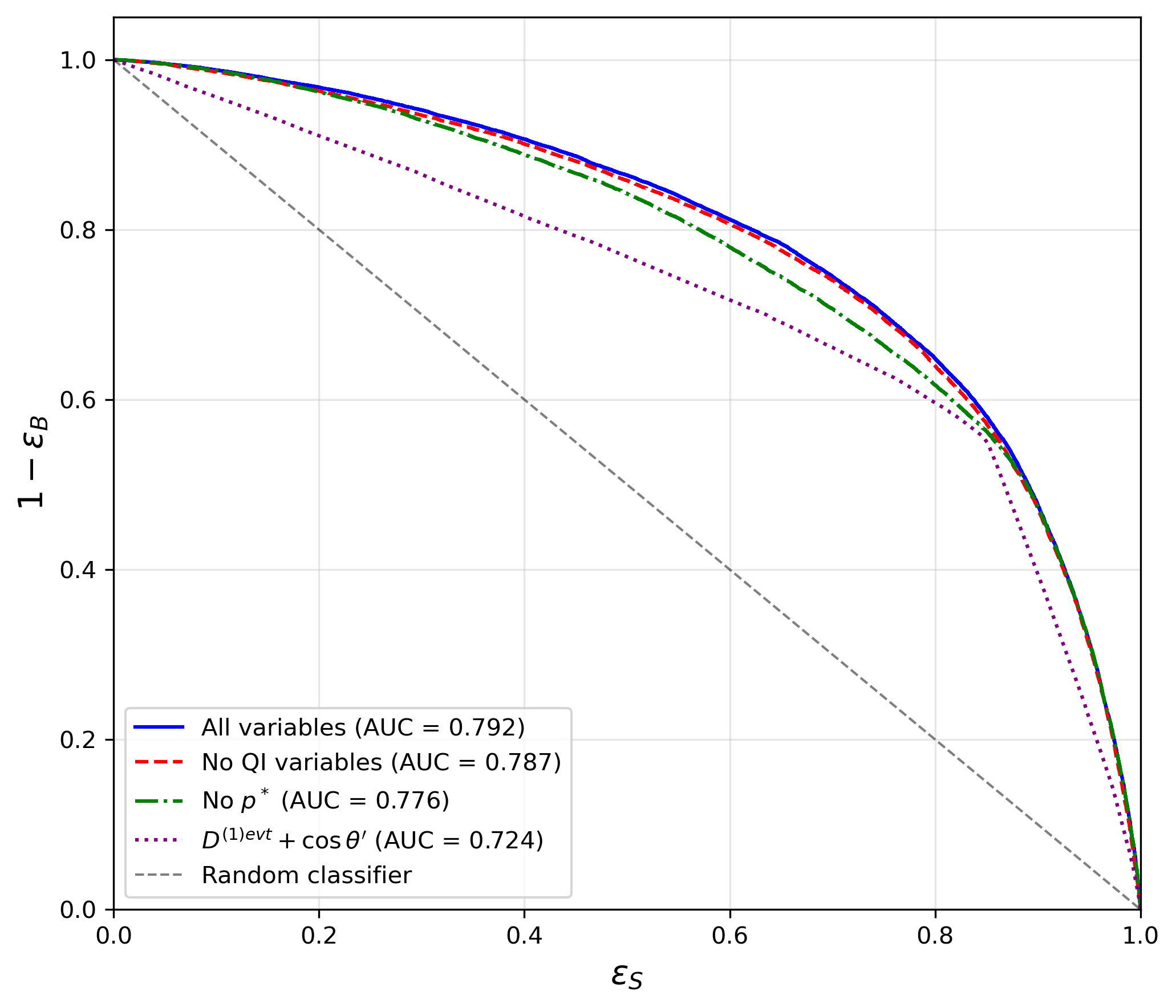}
  \caption{BDT performance for separating the toponium signal from the \ttbar\ background, after the selection cuts of Section~\ref{sec:sim}. \textit{Left panel} -- BDT score distributions shown separately for the training and testing samples.  \textit{Right panel} -- ROC curves and AUC scores for different BDT input configurations including the full set of variables (solid blue), all variables with the exclusion of the quantum-information-inspired observables (dashed red), all variables with the exclusion of the $p^\star$ variable (dash-dotted green) and a configuration using only $D^{(1)\mathrm{evt}}$ and $\cos \theta'$ (dotted magenta). The kink in the two-variable configuration arises from the discrete structure of the tree-based classifier. \label{fig:BDT_performance}}
\end{figure}

For completeness, the left panel in Figure~\ref{fig:BDT_performance} shows the BDT score distributions for the signal and the background. The agreement between training and testing samples validates our procedure and illustrates that no significant overfitting is observed. To assess the robustness of the multivariate strategy, we compare in the right panel of the figure the ROC curves obtained for different choices of input observables. The reference configuration uses the full set of variables introduced in Section~\ref{sec:observables} (solid blue). We then consider reduced configurations in which either the quantum-information-inspired observables are removed (dashed red) or the kinematic variable $p^\star$ is excluded (dash-dotted green), or in which only the spin-correlation observables $D^{(1)\mathrm{evt}}$ and $\cos\theta'$ are used as inputs (dotted magenta). The usage of the full set of observables provides the best overall performance, demonstrating that the different classes of variables contribute with complementary information. Removing the quantum-information-inspired observables results in only a mild degradation in the AUC score, indicating that although these variables do not dominate the discrimination individually, they provide additional sensitivity when combined with the kinematic and angular observables. By contrast, excluding $p^\star$ leads to a more visible loss of performance, confirming its strong sensitivity to near-threshold production dynamics, such a behaviour being consistent with the feature-importance ranking shown in the bottom row of Figure~\ref{fig:BDT_summary} where $p^\star$ appears among the leading discriminants in the low-momentum region. Finally, a classifier based solely on the spin-correlation observables $D^{(1)\mathrm{evt}}$ and $\cos\theta'$ exhibits substantially reduced separation power. This confirms that, while these quantities have a clear physical interpretation in terms of the structure of the spin-density matrix, they are not sufficient on their own to achieve competitive discrimination between toponium and continuum \ttbar\ production. Overall, this comparison demonstrates that the optimal sensitivity is only achieved by combining angular, kinematic and spin-sensitive observables within a global framework. While the results focus so far on statistical discrimination, potential systematic uncertainties such as those associated with detector effects, reconstruction or the modelling of the kinematic and spin-sensitive observables are expected to affect the absolute significance. The relative performance ranking of the different BDT configurations is however expected to be robust to those effects. All these findings are summarised in Table~\ref{tab:bdt_perf_pt35} in which we collect the BDT performance at the optimal working point for the various choices of input variables.  

\begin{table}\renewcommand{\arraystretch}{1.2}\setlength{\tabcolsep}{16pt}
\centering \begin{tabular}{lcccccc}
\toprule 
& ${\varepsilon_S}$ 
& ${\varepsilon_B}$ 
& ${S/\sqrt{B}}$ 
& ${S/\sqrt{S+B}}$ 
& ${S/\sqrt{S+B+\sigma_S^2}}$ 
& {BDT score cut} \\
\midrule
(a) & 0.665     & 0.228 & 37.619 & 37.202 & 4.955 & 0.835 \\
(b) & 0.643     & 0.219 & 37.092 & 36.680 & 4.954 & 0.844 \\
(c) & 0.792     & 0.375 & 34.944 & 34.663 & 4.949 & 0.777 \\
(d) & 0.830     & 0.428 & 34.249 & 33.996 & 4.947 & 0.797 \\
\bottomrule
\end{tabular}
\caption{Performance metrics at the optimal working point for various BDT configurations after selection cuts: (a) all variables described in Section~\ref{sec:observables}; (b) all variables excluding the quantum-information-inspired ones; (c) all variables excluding $p^\star$; (d) using only $D^{(1)\mathrm{evt}}$ and $\cos \theta'$. The signal uncertainty is assumed to be $\sigma_S = 0.2\, S$, and the corresponding optimal BDT score cut for each configuration is indicated.}
\vspace{0.7mm}
\label{tab:bdt_perf_pt35}
\end{table}

\section{Summary and outlook}\label{sec:conclusion}

In this work, we have investigated the impact of toponium formation on spin observables and quantum information properties of dileptonic \ttbar production at the LHC at $\sqrt{s} = 13$~TeV. Treating the \ttbar\ system as a mixed two-qubit state, we have reconstructed the spin density matrix and studied a broad set of observables, including spin correlation coefficients, angular observables and quantities inspired by quantum information theory.

We have first presented results defined over the full event sample, enabling quantum tomography to extract the Fano coefficients of the spin density matrix. Toponium formation induces a characteristic modification of the structure of the matrix: its diagonal components are significantly enhanced, each approaching $-1$. This reflects the bound-state nature of the near-threshold \ttbar system which is produced as a coherent two-particle state with well-defined quantum numbers. Consequently, derived observables such as the $D^{(i)}$ coefficients, the concurrence, the normalised purity, the logarithmic negativity, magic and the trace distance show pronounced differences. Even after accounting for the relative production rates and event selection, these observables still provide a sensitive probe of the quantum correlations induced by toponium effects. We have additionally studied analogous event-by-event variables. Kinematic and angular observables probing the global \ttbar geometry, such as the top-quark momentum in the \ttbar system $p^\star$ and the angular separations $\Delta R$ and $\Delta\phi$ between the leptons resulting from the top and antitop decays, exhibit the strongest single-variable discriminating power. Variables constructed solely from reconstructed top-quark kinematics also retain significant sensitivity to toponium effects. In contrast, observables directly related to the spin correlations such as $D^{(1)\mathrm{evt}}$ and $\cos\theta'$ provide limited discrimination individually, despite their clear physical interpretation.

By combining all observables in a multivariate framework, we demonstrate that optimal sensitivity to toponium formation is achieved when angular, kinematic and quantum-information-inspired variables are exploited simultaneously. Excluding the quantum information observables leads to only a mild reduction in performance, indicating that while these variables are not dominant alone they contribute with complementary information. Omitting the kinematic variable $p^\star$ results in a more noticeable loss of sensitivity, highlighting its central role in highlighting toponium effects. Overall, our results show that quantum-information-inspired observables provide a consistent and physically transparent framework to characterise the spin structure of near-threshold \ttbar production. While their standalone discriminating power is limited, they enhance the interpretability of the analysis and improve multivariate sensitivity when combined with conventional angular and kinematic observables. Future studies incorporating detector effects, systematic uncertainties and potential interference from beyond-the-Standard-Model contributions will be essential to fully assess the experimental reach of this approach and its applicability to precision top-quark studies at the LHC.

\section*{Acknowledgements}

BF is supported in part by Grant ANR-21-CE31-0013 (Project DMwithLLPatLHC) from the French \emph{Agence Nationale de la Recherche} (ANR). AO acknowledges financial support by CF-UM-UP through the Strategic Fundings UIDB/04650/2020, UIDP/04650/2020, UID/PRR/04650/2025 and UID/04650/2025 from \emph{Funda\c{c}\~ao para a Ci\^encia e Tecnologia} (FCT). EC is supported by the Funda\c{c}\~ao para a Ci\^encia e a Tecnologia (FCT) under the contract PRT/BD/154189/2022. M.C.N.F. was supported by the PSC-CUNY Award 68031-00 56. MJW is supported by the Australian Research Council grants CE200100008 and DP220100007.

\bibliographystyle{JHEP}
\bibliography{references}

\end{document}